\documentclass{aa} 
\usepackage{graphicx}
\usepackage[varg]{txfonts}
\usepackage{subfig}
\usepackage[labelfont=bf]{caption}
\usepackage{graphicx}
\usepackage{longtable,booktabs}
\usepackage{pdfpages}
\usepackage{import}
\usepackage[maxfloats=256]{morefloats}
\newcommand{\kms}{~km~s$^{-1}$}
\newcommand{\msun}{M$_\odot$}
\newcommand{\barolo}{$^{\mathrm{3D}}$\textsc{Barolo} }
\newcommand{\halpha}{\ion{H}{$\alpha$}}
\newcommand{\hbeta}{\ion{H}{$\beta$}}
\newcommand{\hi}{\ion{H}{i}}
\newcommand{\hii}{\ion{H}{ii}}
\newcommand{\nii}{[\ion{N}{ii}]}
\newcommand{\sii}{[\ion{S}{ii}]}

\defcitealias{myGASPOnlineAtlas}{Atlas}

\graphicspath{{./}}

\usepackage[]{hyperref}
\hypersetup{colorlinks=true,linkcolor=cyan,citecolor=cyan,urlcolor=magenta}

\begin{document}

\title{Impact of the large scale environment on the baryonic scaling relations of disc galaxies: Insights from the GASP survey}

\authorrunning{C. Bacchini et al.}
\titlerunning{Baryonic scaling relations using GASP}
\author{C. Bacchini~\inst{\ref{inst:dark}}, 
		A. Marasco~\inst{\ref{inst:inaf_pd}},
		B. Vulcani~\inst{\ref{inst:inaf_pd}},
		A. Moretti~\inst{\ref{inst:inaf_pd}},
		J. Fritz~\inst{\ref{inst:morelia}},
		R. Paladino~\inst{\ref{inst:eso},\ref{inst:inaf_bo}},
		B. M. Poggianti~\inst{\ref{inst:inaf_pd}},
		M. Mingozzi~\inst{\ref{inst:esaaura}},
		P. E. Mancera Pi$\mathrm{\tilde{n}}$a~\inst{\ref{inst:leiden}}, 
		F. Rizzo~\inst{\ref{inst:kapt}}
			}

\institute{
DARK, Niels Bohr Institute, University of Copenhagen, Jagtvej 155, 2200 Copenhagen, Denmark \label{inst:dark}\\
\email{cecilia.bacchini@eso.org}
\and
INAF - Osservatorio Astronomico di Padova, vicolo dell'Osservatorio 5, 35122 Padova, Italy \label{inst:inaf_pd}
\and
Instituto de Radioastronomía y Astrofísica, UNAM Campus Morelia, A.P. 3–72, C.P. 58089, Mexico \label{inst:morelia}
\and 
European Southern Observatory, Karl-Schwarzschild-Strasse 2, 85748 Garching, Germany \label{inst:eso}
\and
INAF - Istituto di Radioastronomia, Via P. Gobetti 101, 40129 Bologna, Italy \label{inst:inaf_bo}
\and
AURA for ESA, Space Telescope Science Institute, 3700 San Martin Drive, Baltimore, MD 21218, USA \label{inst:esaaura}
\and
Leiden Observatory, Leiden University, P.O. Box 9513, 2300 RA, Leiden, The Netherlands \label{inst:leiden}
\and 
Kapteyn Astronomical Institute, University of Groningen, Landleven 12, 9747 AD, Groningen, The Netherlands \label{inst:kapt}
}

\date{Draft xx; Received xx / Accepted xx.}

\abstract{
The Tully--Fisher relation (TFR) and the specific angular momentum--mass relation (SAMR) link the baryonic content of a disc to its kinematics and dynamics. 
Hence, these scaling relations should bear the imprints of any process removing baryons or altering their kinematics, such as those acting in clusters of galaxies. 
We test this using the MUSE survey "GAs Stripping Phenomena in galaxies" (GASP), which targets field, group, and cluster members at $z \approx 0.05$ and various stages of ram pressure stripping. 
For the 81 discs with favourable inclinations, we measure the distribution and kinematics of stars and ionised gas, and their specific angular momenta. 
Within their scatter, the stellar and baryonic relations show no significant dependence on the large scale environment: the TFRs have slopes $\alpha \approx 3.6$ and $\approx 3.5$ for any kinematic tracer, and the SAMRs have $\alpha \approx 0.65$--$0.75$, as predicted by tidal torque theory. 
The same holds for the ten S0-S0/as in our sample, consistent with the "faded spiral" formation scenario. 
Only the gas SAMR shows an environmental signature, with ram pressure stripped galaxies lying $\approx 0.16$~dex below unperturbed ones. 
This offset is mostly due to their enhanced star formation rate, and thus higher gas mass when inferred through the Schmidt-Kennicutt law, rather than to kinematic perturbations or disc truncation. 
Our results suggest that, within the area probed by MUSE ($\lesssim6$ disc scale lengths), these discs have preserved the angular momentum acquired at formation not only from cluster processes and pre-processing in filaments, but also from mergers and ejective feedback acting in any environment. 
}

\keywords{Galaxies: cluster -- Galaxies: elliptical and lenticular, cD -- Galaxies: fundamental parameters -- Galaxies: kinematics and dynamics -- Galaxies: spirals -- ISM: kinematics and dynamics}

\maketitle

\section{Introduction}\label{sec:Intro}
The Tully--Fisher relation (TFR) and the specific angular momentum--mass relation (SAMR) link key properties of disc galaxies: mass ($M$), circular speed ($V_\mathrm{circ}$), and angular momentum ($J$). 
The stellar TFR reads $M_\star\footnote{Subscripts $\star$, gas, \hi, H$_2$, \hii, and bar indicate, respectively, stars, total cold gas, atomic gas, molecular gas, ionised gas, and baryons.} \propto V_\mathrm{circ}^\alpha$, with $\alpha \approx 3.5-5$ and scatter as low as 6\%, depending on how $M_\star$ and $V_\mathrm{circ}$ are defined \citep[e.g.][]{1977TullyFisher,2001Verheijen,2017Ponomareva,2018Ponomareva,2016Lelli_a,2025Marasco}. 
Traditionally, the TFR is built from the atomic hydrogen (\hi) rotation curves, which trace well $V_\mathrm{circ}$ except for the least massive galaxies \citep{2017Iorio}. 
Replacing $M_\star$ with the baryonic mass $M_\mathrm{bar}=M_\star+M_\mathrm{gas}$ gives a tighter relation \citep{2000McGaugh}: $M_\mathrm{bar}\propto V_\mathrm{circ}^\alpha$, with $\alpha\approx3.6-4$ and orthogonal intrinsic scatter $\delta_\perp\approx0.03$ when $V_\mathrm{circ}$ is computed from the flat part of the \hi\ rotation curve \citep[e.g.][]{2019Lelli}. 
The baryonic TFR is a key benchmark for galaxy evolution models and alternative theories to $\Lambda$ cold dark matter ($\Lambda$CDM) cosmology \citep[e.g.][]{2000McGaugh,2016Lelli_a,2019ManceraPina_b,2024ManceraPina}.

The SAMR was first proposed by \cite{1983Fall}, who argued that baryons initially inherit the same $j$ as their dark matter (DM) haloes through tidal torques from neighbouring overdensities \citep{1969Peebles,1980Fall}. 
The $\Lambda$CDM paradigm predicts $j_\mathrm{DM} \propto M_\mathrm{DM}^{2/3}$ for DM haloes. 
The Fall relation, $j_\star \propto M_\star^\alpha$ with $\alpha \approx 2/3$, follows if $j_\mathrm{bar}$ is conserved as baryons collapse into a star-forming disc. 
Empirical determinations of the SAMR for stars, molecular gas, and total baryons in nearby galaxies find $\alpha \approx 0.5-0.7$, suggesting that their $j$ is largely conserved, whereas the SAMR for \hi\ alone has $\alpha \approx 1$ \citep[e.g.][]{2012RomanowskyFall,2018Posti,2021ManceraPina,2023DiTeodoro,2025Geesink}.

These scaling relations reflect the complex interplay of cooling, star formation, and feedback from stars and active galactic nuclei (AGN). 
AGN feedback, for example, can preferentially remove low $j$ gas from galaxy centres: if the outflow escapes the galactic potential, $M_\mathrm{gas}$ decreases while $j_\mathrm{gas}$ increases, and stars formed afterwards inherit this signature \citep[e.g.][]{2012RomanowskyFall}. 
Besides internal processes like feedback, the large scale environment can also leave imprints on $V_\mathrm{rot}$ (or $V_\mathrm{circ}$), $j$, and $M$.
Galaxies in clusters and dense groups undergo both hydrodynamical and gravitational processes \citep{2021Cortese}, which leave different signatures on the scaling relations.
The dominant hydrodynamical mechanism is likely the ram pressure exerted by the intra-cluster medium \citep[ICM;][]{1972GunnGott,2022Boselli}, which strips gas from the interstellar medium (ISM), most efficiently in the outer disc where the gravitational restoring force is weak. 
Because most $j$ resides at large galactocentric radii, ram pressure is expected to reduce $j_\mathrm{gas}$ more efficiently than $M_\mathrm{gas}$ \citep{1983Fall,2012RomanowskyFall}, displacing galaxies below the gas SAMR while leaving the galaxy potential, and hence $V_\mathrm{circ}$, mostly unchanged.
Stars are not directly affected \citep[although see][]{2012Smith}, so $M_\star$ and $j_\star$ are preserved if stripping occurs after most of $M_\star$ has formed. 
Instead, early and/or prolonged gas loss over several Gyr may result in lower $M_\star$ and $j_\star$.
Ram pressure can also effectively remove the circumgalactic medium, the main reservoir for gas accretion \citep[e.g.][]{2017Fraternali,2023Afruni}; the remaining gas in the ISM is then consumed by star formation processes over timescales of a few Gyr \citep{2021Cortese,2022Boselli}, without altering $M_\star$ and $j_\star$.

Gravitational processes act instead on both gas and stars.
Tidal forces from galaxy--galaxy interactions or the cluster DM halo can remove the CGM and the ISM, and also heat the disc dynamically, raising the velocity dispersion and lowering $V_\mathrm{rot}$, and hence $j_\mathrm{gas}$ and $j_\star$ \citep[e.g.][]{2011Bekki}.
Major mergers can destroy stellar discs, turning a rotationally supported galaxy into a dispersion dominated one with lower $j_\star$ \citep{2014Naab,2018Lagos}.
Although a rotating disc may reform from the remaining or newly accreted gas, mergers are generally expected to reduce $j$ by $\approx30$--40\% while increasing $M$ \citep[][and \citealt{2018Lagos} for a discussion of wet and dry mergers]{2017Lagos}.
Cluster galaxies may also have been pre-processed by interactions and ram pressure in filaments and groups before infall \citep[e.g.][]{2026Rhee,2026Vulcani}, so their present-day location on the scaling relations reflects both phases.

Observationally, how the environmental processes affect disc scaling relations is not fully settled.
Many studies of the TFRs and SAMRs rely on \hi\ to trace the baryon kinematics, by now covering $\approx 6$~dex in $M_\mathrm{bar}$ \citep[e.g.][]{2025ManceraPina}. 
In fact, \hi\ is the best tracer of the disc outskirts, where most of $j$ resides and rotation curves flatten, minimising the scatter of the TFRs \citep{2001Verheijen,2019Lelli} and SAMRs \citep{2018Posti}. 
Moreover, \hi\ is the main ISM component of star forming galaxies and gives the most accurate gas masses, as its mass-to-light ratio is much less uncertain compared to CO for H$_2$ or \halpha\ for ionised gas. 
However, interferometric \hi\ observations at high spatial resolution and signal-to-noise ratio (S/N) become prohibitively expensive beyond $\approx100$~Mpc, so many studies of scaling relations measure $V_\mathrm{circ}$ from the integrated \hi\ profile \citep{2022Hardwick,2023Catinella,2023Gogate}, which is not ideal because it is biased towards the velocities of dense gas in galaxy centres \citep{2019Lelli}. 
Environmental studies based on spatially-resolved \hi\ rotation curves are therefore limited to Virgo \citep{1988Guhathakurta} and Ursa Major \citep{2001Verheijen}, although Square Kilometre Array (SKA) precursors are improving the situation \citep[e.g.][]{2025Ndaliso}. 
Recently, \citet{2024Sorgho} have found that isolated galaxies have higher $j_\mathrm{bar}$ than those with close companions, without linking this difference to the large scale structure. 
The gas, which is the component most directly affected by ram pressure, remains the least explored: studies of clusters have probed $j_\mathrm{bar}$ and the atomic gas fraction \citep[e.g.][]{2020Li_b,2020Murugeshan}, but not the gas SAMR itself.

With respect to \hi\ surveys, optical integral field spectroscopic surveys reach high spatial resolution on much larger samples and measure the stellar kinematics directly, avoiding the empirical corrections or assumptions on the stellar velocity dispersion needed to infer the stellar $V_\mathrm{rot}$, and hence $j_\star$, from $V_\mathrm{rot,\hi}$ \citep[e.g.][]{2019Marasco,2026ManceraPina}. 
However, the limited field of view (FOV) of integral field units (IFU) usually covers $\lesssim2$--3 effective radii \citep{2015Bundy}.
This is suboptimal to measure the flat part of rotation curves and the bulk of $j$, and might influence the resulting scaling relations in a non trivial way \citep[e.g.][]{2019Lelli,2023Catinella}.
Notwithstanding these issues, optical studies found that the TFR in dense environments is indistinguishable from that of the field, both in Hickson compact groups \citep{2013TorresFlores} and in the Mapping Nearby Galaxies at Apache Point Observatory survey \citep[MaNGA;][]{2024Ristea}.
Analyses of the Sydney-AAO Multi-object Integral field (SAMI) survey are based on the stellar spin parameter $\lambda_\star$ \citep{2007Emsellem}, which quantifies the rotational support rather than $j_\star$ \citep{2018Rizzo}, and conclude that the environment affects it mostly indirectly via quenching, so older galaxies have lower spin mostly because stars were born dynamically hotter at high $z$ or secularly heated \citep{2019Cortese,2024Croom}, in line with simulations finding $j_\star$ is preserved upon infall \citep{2021Mostoghiu}.

This work investigates the impact of environment on the TFRs and SAMRs using the IFU survey ``GAs Stripping Phenomena in galaxies'' \citep[GASP;][]{2017Poggianti,2025Poggianti}, carried out with the Multi Unit Spectroscopic Explorer (MUSE). 
Its advantage over other IFU surveys is the large FOV and sensitivity of MUSE, which reach $\gtrsim10\,R_\mathrm{e}$ while keeping good spatial ($\approx1$~kpc) and spectral ($\sigma_\mathrm{MUSE}\approx50$~\kms) resolution \citep{2017Poggianti}. 
GASP also includes star-forming and passive discs in clusters, groups, and the field, and is thus well suited to environmental studies of disc scaling relations \citep[e.g.][]{2020Vulcani_b}.
Our aim is to investigate and quantify the effect of the environment, and of ram pressure in particular, on the stellar and baryonic TFRs and the SAMRs for stars, gas, and baryons.
To minimise the systematics, we adopt consolidated methods from \hi\ based studies, since the environmental offsets may be comparable to the intrinsic scatter of the relations. 
This, combined with the larger radial coverage provided by our MUSE data, are the main improvements with respect to IFU based studies. 
We also provide a public galaxy atlas and the first characterisation of the TFRs and SAMRs for the GASP sample. 
Section~\ref{sec:SampleAndData} describes observations, galaxy sample, and data products. 
Section~\ref{sec:BaryDist} derives the galaxy geometry and the distributions and masses of gas and stars. 
Section~\ref{sec:BaryKin} presents the kinematic modelling and circular velocity curves. 
Section~\ref{sec:results} analyses the dependence on environment of baryon kinematics and scaling laws; these results are discussed in Sect.~\ref{sec:discussion}. 
Section~\ref{sec:conclusions} summarises our conclusions.

\section{Parent sample and observations}\label{sec:SampleAndData}
This section describes the GASP sample, observations, and data products (Sects.~\ref{sec:GASP} and~\ref{sec:HICO}), as well as the classifications for ram pressure stripping, AGN, and morphology (Sect.~\ref{sec:class}).

\subsection{MUSE data from ``GASP''}\label{sec:GASP}
To maximize the sample size, we consider all galaxies observed by the GASP survey \citep{2017Poggianti,2025Poggianti}, which targeted 114 galaxies. 
Of these, 94 are drawn from the catalogue of stripped galaxies from \cite{2016Poggianti}, based on visual classification of $B$ and $V$ band images and including galaxies with morphological signatures of gas stripping across different environments. 
The remaining 20 galaxies constitute a control sample of undisturbed systems selected from the WINGS/OmegaWINGS surveys \citep{2006Fasano,2014Moretti,2015Gullieuszik,2017Moretti} and the PM2GC catalogue \citep{2011Calvi}. 
We also consider three galaxies (with suffix "b") within the MUSE FOV of JO119, JO157, and P17048 having similar $z$ to the targets, and JW105b at $z\approx 0.3762$. 
In total, our parent sample contains 118 galaxies, listed in Table~\ref{tab:sample}.
The MUSE data cover $4800 \lesssim \lambda/\text{\AA} \lesssim 9300$ with resolving power $1770 < \mathrm{R}<3590$ and were collected with seeing of $\approx 1$\arcsec. 
The spaxel volume is 0.2\arcsec$\times$0.2\arcsec$\times1.25$\AA. 

This work builds on several data products from the reduced MUSE observations \citep{2017Poggianti,2025Poggianti} and shown in the online \citetalias{myGASPOnlineAtlas}. 
$I$~band images were obtained by \cite{2020Franchetto} from the integrated MUSE datacubes using the Cousins $I$~band filter response curve. 
We use these images to characterise the stellar distribution and substructures such as bulges and bars (Sec.~\ref{sec:StarDist}), which are dominated by evolved ($> 1$~Gyr) populations and thus brighter in $I$~band than at shorter $\lambda$ \citep[e.g.][]{2000Knapen}. 
The stellar velocity fields and velocity dispersion maps were extracted from the MUSE datacube \citep[see][]{2017Poggianti} using the penalized pixel-fitting code \textsc{pPXF} \citep{2004Cappellari,2017Cappellari}. 
Spurious sources (e.g. background stars and galaxies) were masked out and the spaxels Voronoi-binned to $\mathrm{S/N}>10$. 
Spectra were fitted with single stellar population templates from \cite{2010Vazdekis}, assuming a single kinematic component. 
This may oversimplify galaxies with a prominent bulge or bar, which are kinematically distinct from the disc \cite[e.g.][]{2017Tabor,2017Cappellari,2018Rizzo}, but our galaxies are mostly disc dominated, so bulge contamination is usually secondary. 

The stellar mass maps were obtained using the spectral synthesis code \textsc{Sinopsis} \citep{2011Fritz,2017Fritz} and presented in \cite{2018Vulcani,2020Vulcani_a,2021Vulcani}. 
\textsc{Sinopsis} decomposes galaxy spectra into stellar population models, fitting the continuum and absorption and emission lines, and provides maps of the population properties (e.g. mass, age) and continuum subtracted datacubes with emission lines only. 
From these we build the H$\alpha$ and H$\beta$ velocity datacubes used for the ionised gas kinematics (Sec.~\ref{sec:GasKin}), following the procedure of \citet{2023Marasco}: a third-order polynomial is fitted in a 120~\AA-wide window around each line and then subtracted to remove the residual continuum. 
This reduces spurious wings in the line profiles, which may mimic outflows or inflate the velocity dispersion. 
For some galaxies, the doublet \nii$\lambda\lambda6548,6583$ is so broadened by the galactic rotation and/or outflows that it ``contaminates'' the \halpha\ datacube. 
We remove it by masking the \halpha\ datacubes before the kinematic modelling, although residual \nii\ emission might remain in the innermost regions of a few galaxies. 
Since \nii\ is much fainter than \halpha, this is not a big issue for our analysis (see Sect.~\ref{sec:GasKin}). 
The final H$\alpha$ and H$\beta$ datacubes have channel separations of $\approx 53 - 55$~\kms and $\approx 67 - 75$~\kms, respectively.

\subsection{HI and CO emission observations}\label{sec:HICO}
We constrain the cold gas content using follow-up programs targeting the \hi\ 21-cm emission with the Very Large Array and MeerKAT, and the CO emission with the Atacama Large Millimeter Array (ALMA) and the Atacama Pathfinder Experiment (APEX). 
\hi\ and APEX data have too coarse resolution to resolve the gas distribution and kinematics, but give the \hi\ ($M_\mathrm{\hi}$) and H$_2$ ($M_\mathrm{H_2}$) masses for ten and nine galaxies, respectively (see Table~\ref{tab:sample}). 
The ALMA data resolve the discs ($\approx 1$~kpc) but cover only four galaxies \citep{2020MorettiLetter,2023Bacchini}. 
In Sect.~\ref{sec:GasDist}, we infer the cold gas distribution from \halpha\ using $M_\mathrm{\hi}$ (multiplied by 1.36 to account for the Helium mass) and $M_\mathrm{H_2}$ (assuming the Milky Way CO-to-H$_2$ conversion factor and a factor 2 uncertainty) from previous studies \citep{2019Ramatsoku,2020Ramatsoku,2020Deb,2022Deb,2023Moretti}. 

\subsection{Classifications from previous studies}\label{sec:class}
To search for environmental effects we divide the sample into cluster and field galaxies, the latter including group members, treated as low density environments.
We adopt the cluster member classification of \cite{2016Poggianti,2025Poggianti}, while group and field galaxies are identified by \cite{2018Vulcani,2021Vulcani}; Table~\ref{tab:sample} reports the environment for each galaxy.
We distinguish between early-type (passive) discs and late-type (star forming) ones using the morphological classifications based on $B$ and $V$ photometric images (\citealt{2014Calvi}; \citealt{2015Fasano}; \citealt{2022Vulcani} for field galaxies; \citealt{2021Vulcani,2022Vulcani}; \citealt{2025Poggianti} for cluster members). 
We consider together passive discs: "classic" lenticulars (S0s) without spiral arms and with bulges, and those sharing properties with Sa (hereafter S0/a), such as spiral arms. 
Based on MUSE data, we revise the type of four galaxies (see Appendix~\ref{app:atlas}).
Table~\ref{tab:sample} reports the morphological types for all galaxies.

For cluster members we adopt the JType classification of \cite{2025Poggianti}, which grades the degree of ram pressure stripping. 
Galaxies with unilateral \halpha\ tails are classified as "strong" or "extreme" stripping depending on whether the tail is moderately or significantly more extended than the stellar disc diameter. 
"Weak" and "mild" stripping classes show some \halpha\ emission outside the stellar disc, while "truncated" discs have \halpha\ much less extended than the stellar disc. 
Galaxies with no \halpha\ are "fully stripped" and are all passive disks, mostly post-starburst. 
For group and field galaxies, we adopt the classification by \cite{2021Vulcani}. 
They combine stellar and ionised gas kinematics, stellar population properties, and BPT diagrams to identify the mechanisms at work in low-density environments: interactions, mergers, ram pressure in groups, cosmic web stripping (ram pressure in filaments), star formation enhancement, gas accretion, and starvation. 
Most field galaxies are affected by none or other processes, but four (P20159, P59597, P96244, and P18060) have \halpha\ tails due to ram pressure or cosmic web stripping. 
To homogenise the classification with that of cluster members, we assign them JType=1 (strong stripping) and JType=3 to P5055, which hosts a truncated \halpha\ disc \citep{2021Vulcani}. 
JTypes for the parent sample are in Table~\ref{tab:sample}. 
We use catalogues of AGN hosts in clusters \citep{2017PoggiantiNat,2022Peluso} and in low density environments \citep{2021Vulcani}, identified from optical line ratios and Baldwin-Phillips-Terlevich (BPT; \citealt{1981Baldwin}) diagrams. 

In summary, the parent sample has 118 galaxies, 76 in clusters and 42 in low density environments. 
13 galaxies are S0 or S0/a (ten in clusters, one in the field, and two in groups); 14 galaxies, of which 12 are in clusters, host an AGN. 
Among cluster members, 21 show no to mild signs of ram pressure stripping, 36 are strongly stripped or jellyfish galaxies (extreme stripping), four have truncated \halpha\ discs, nine are fully stripped, and six are mergers or unknown cases \citep{2025Poggianti}. 
In groups, four galaxies are undergoing ram pressure stripping, one is a truncated disc. 
Excluding JW105b, the remaining 37 galaxies are affected by none or other mechanisms not investigated here. 
Our parent sample carries an inclination bias, likely because the visual detection of extra-planar tails is easier in highly inclined discs. 
In fact, the median inclination increases from $\approx 30\deg$ for the undisturbed galaxies to $\approx 60-75\deg$ for the most stripped ones, which motivates the inclination cuts adopted in Sect.~\ref{sec:results}. 


\section{Distribution of baryonic components}\label{sec:BaryDist}
In this section, we measure the stellar geometry (Sect.~\ref{sec:Geometry}), and the mass and distribution of stars (Sec.~\ref{sec:StarDist}) and gas (Sec.~\ref{sec:GasDist}). 

\subsection{Galaxy geometry}\label{sec:Geometry}
Measuring the disc inclination ($i$) is crucial to obtain its $V_\mathrm{rot}$ from the line-of-sight (LOS) velocities ($V_\mathrm{LOS}$). 
Moreover, determining the extent of the stellar disc ($R_\mathrm{max}$) is key to compute the total $j_\star$, mostly locked at $R > 2-3 R_\mathrm{e}$ \citep{2012RomanowskyFall}. 
\cite{2020Franchetto} measure the stellar disc geometry via isophotal analysis on the $I$~band images, which are more axisymmetric than stellar mass maps for galaxies with extra-planar star formation (e.g. jellyfish galaxies). 
We refine their isophotal analysis by computing the background with the more sophisticated procedure of \cite{2023Marasco}, detailed in Appendix~\ref{app:isophotes}. 
We use the Python package \textsc{photutils} to fit the background-subtracted images with a set of elliptical isophotes defined by the centre coordinates ($x_0,y_0$), the position angle $\phi$, and the ellipticity $\epsilon = 1 -b/a$, with $a$ and $b$ respectively the ellipse major and minor axis. 
The ellipse inclinations are derived from $a/b$, and the disc $i$ and $\phi$ are the medians over the radial range that excludes bars and bulges (see Appendix~\ref{app:isophotes}). 
We then extract the mean intensity $I(R)$ in progressively larger ellipses and define $R_\mathrm{max}$ as the radius where $I(R)$ drops below the noise (Appendix~\ref{app:isophotes}). 
The \citetalias{myGASPOnlineAtlas} shows the $I$ band images and the isophote fitting with its geometrical parameters. 
Our geometry and $I(R)$ agree within the uncertainties with \cite{2020Franchetto}, though our $R_\mathrm{max}$ is slightly larger for some galaxies. 

\subsection{Stellar distributions and masses}\label{sec:StarDist}
We derive the radial profiles of the stellar mass surface density $\Sigma_\star(R)$ using \textsc{Sinopsis} stellar mass maps, dividing them in rings with the geometry from Sect.~\ref{sec:Geometry}. 
In each ring, we compute the mean $\Sigma_\star$ and its uncertainty (see Appendix~\ref{app:errors}). 
Within this region, our $\Sigma_\star(R)$ (corrected for $i$) matches that from \citet{2020Franchetto}. 
Panels C and D in the \citetalias{myGASPOnlineAtlas} show the stellar mass maps and $\Sigma_\star(R)$. 
Appendix~\ref{app:bulge_disc} details how we separate the stellar disc and bulge mass distributions, which are modelled as an exponential and a S\'{e}rsic component, respectively. 
The disc ($M_{\star,d}$) and bulge ($M_{\star,b}$) masses are computed using Eq.~2.166 of \cite{2008BinneyTremaine} and Eq.~A2 of \cite{2005Terzic}, extrapolating the models to $R_\mathrm{max}$, with uncertainties from Monte Carlo realisations of the best-fit models. 
The bulge fraction is $\mathrm{B/T} = M_{\star,b}/M_\star $, where $M_\star = M_{\star,d} + M_{\star,b}$. 
Our total $M_\star$ are consistent within the uncertainties with previous studies \citep[see e.g.][]{2025Poggianti}. 
Panel D in the \citetalias{myGASPOnlineAtlas} shows the bulge and disc models, their parameters, $\Upsilon^I$, masses, and $B/T$.

\subsection{Gas distributions and masses}\label{sec:GasDist}
Since most of our galaxies lack cold gas observations, we infer $M_\mathrm{gas}$ from the star formation rate (SFR) surface density ($\Sigma_\mathrm{SFR}$) traced by \halpha, inverting the Schmidt-Kennicutt (SK) law. 
\textsc{Sinopsis} yields maps of the SFR over the last 20~Myr, but we prefer \halpha\ because it reaches larger radii, being brighter than the stellar continuum and unaffected by the S/N cut of the SED fitting. 
For star-forming galaxies, we derive the azimuthally-averaged surface brightness profiles of H$\alpha$ ($\Sigma_\mathrm{H\alpha}$) and H$\beta$ ($\Sigma_\mathrm{H\beta}$) from the emission line datacubes. 
We identify the regions with contamination by the AGN and \nii\ emission by visually inspecting the position-velocity diagram (PVD) along the major axis from H$\alpha$ datacubes (see for instance JO204 in \citetalias{myGASPOnlineAtlas}). 
AGN contamination is limited to the innermost radii ($R \approx 0.5-4$~kpc) and broadly coincides with the AGN-ionisation zones identified from BPT maps \citep{2017Poggianti,2019Radovich,2022Peluso}. 
We exclude them with a cut in radius on $\Sigma_\mathrm{H\alpha}$, for consistency with the kinematic modelling (Sec.~\ref{sec:GasKin}). 
Contamination beyond the centre is minor, being only in a few pixels, and the azimuthal averaging mitigates it further. 

We correct $\Sigma_\mathrm{H\alpha}(R)$ for dust extinction using the Balmer decrement method, assuming an intrinsic H$\alpha$-to-H$\beta$ flux ratio of 2.86, the extinction law from \cite{1989Cardelli}, and $R_V=3.1$ (with $\kappa_\mathrm{H\alpha} \approx 2.54$ and $\kappa_\mathrm{H\beta} \approx 3.61$). 
The dust extinction correction is applied to all galaxies except JO144 and JO194, wherever $\Sigma_\mathrm{H\alpha}(R)/\Sigma_\mathrm{H\beta}(R)\geq 2.86$. 
For JO144 and JO194, the correction yields central gas densities ($\Sigma_\mathrm{gas} \gtrsim 10^3$ \msun~pc$^{-2}$) that exceed $\Sigma_\star$ and are therefore likely overestimated; thus, the dust correction is not applied in these two cases. 
We verified that this choice does not affect our main conclusions, although JO144 would become a strong outlier in the scaling relations. 
We prefer applying the dust extinction correction on radial profiles rather than maps because, although local extinction peaks are smoothed out, the azimuthal averaging gives higher S/N than pixel-by-pixel measurements. 
This improves the Balmer decrement, particularly where \hbeta\ is faint, and yields $\Sigma_\mathrm{gas}(R)$ out to larger radii. 
This is important to compute $j_\mathrm{gas}$ over the largest possible area of the star-forming disc. 
In the high S/N regions, $\Sigma_\mathrm{H\alpha}(R)/\Sigma_\mathrm{H\beta}(R)$ matches the azimuthally-averaged profile of the Balmer decrement map. 

From the extinction corrected $\Sigma_\mathrm{H\alpha}(R)$ we derive the \halpha\ luminosity $L_\mathrm{\halpha}$, the integrated $\mathrm{SFR} = 4.6 \times 10^{-42} L_\mathrm{\halpha}$ \citep{1998Kennicutt}, and the profile $\Sigma_\mathrm{SFR}(R)$, assuming a \cite{2003Chabrier} initial mass function (IMF) as in \textsc{Sinopsis} and previous GASP papers; the SFRs agree with previous derivations \citep[e.g.][]{2018Vulcani_b}. 
We then average $\Sigma_\mathrm{SFR}(R)$ over the star-forming disc and convert it into an average $\Sigma_\mathrm{gas}$ by inverting the "global" SK law\footnote{Studies of the resolved SK law, albeit for H$_2$ only, hint at a varying relation in cluster galaxies \citep{2023JimenezDonaire,2022Zabel,2026Moretti}, but it is unknown whether this holds also for the global relation for the total gas.}:
\begin{equation}\label{eq:kslaw}
	\log \left( \frac{\Sigma_\mathrm{SFR}}{\mathrm{M}_\odot \mathrm{yr}^{-1} \mathrm{kpc}^{-2}} \right) = 
	A +
	a
	\log \left(\frac{\Sigma_\mathrm{gas}}{\mathrm{M}_\odot \mathrm{pc}^{-2}} \right)
	\,,
\end{equation}
with the best-fit parameters in \citet{2021Kennicutt} for normal (i.e. non starburst) galaxies, $a = 1.41 \pm 0.087$ and $A = -3.79 \pm 0.08$, where $0.05$~dex has been subtracted from the normalisation to convert from Kroupa to Chabrier IMF. 
The uncertainties on $\Sigma_\mathrm{gas}(R)$ are computed with a Monte Carlo scheme (Appendix~\ref{app:errors}); the central gaps left by the removal of AGN-contaminated radii are filled with the innermost uncontaminated values. 
We include Helium and molecules by multiplying $\Sigma_\mathrm{gas}(R)$ by 1.36 (not included in the best fit of \citealt{2021Kennicutt}). 
$M_\mathrm{gas}$ is computed by integrating $\Sigma_\mathrm{gas}(R)$; S0s and S0/as have no \halpha\ emission from recent star formation, so this procedure gives null $M_\mathrm{gas}$. 
For the 10 galaxies with \hi\ and CO observations, we rescale $\Sigma_\mathrm{gas}(R)$ so that $M_\mathrm{gas}$ matches the sum of the observed M$_\mathrm{\hi}$ and M$_\mathrm{H_2}$. 
These $M_\mathrm{gas}$ are consistent within a factor of $\approx 2$ with the values obtained by inverting the SK law. 
For the four galaxies with resolved CO observations \citep{2020Moretti,2023Bacchini,2026Moretti}, all AGN hosts, our $\Sigma_\mathrm{gas}$ agrees with the observed molecular gas distribution within the uncertainty on the CO-to-H$_2$ conversion factor. 
In Sect.~\ref{sec:AMR}, we discuss how the assumed shape of the SK law influences the SAMR but not our main conclusions. 

\section{Kinematics of baryonic components}\label{sec:BaryKin}
This section describes how we measure the kinematics of stars (Sec.~\ref{sec:StarKin}) and gas (Sec.~\ref{sec:GasKin}). 
For both we use \barolo\footnote{\url{https://editeodoro.github.io/Bbarolo/}} \citep[v.6.1][]{2015DiTeodoro,2021DiTeodoro}, which simulates the observation with a tilted-ring model, consisting of a series of annuli described by geometric and kinematic parameters that vary with the galactocentric distance $R$. 
The geometric parameters are $x_0$, $y_0$, $\phi$, and $i$, while the kinematic ones are the systemic velocity $V_\mathrm{sys}$, $V_\mathrm{rot}$, the velocity dispersion $\sigma$, and the radial velocity in the disc plane $V_\mathrm{rad}$. 
More details on our \barolo\ setup are given in Appendix~\ref{app:kin_setup}. 

Tilted-ring modelling applies to rotating discs, not to galaxies with chaotic kinematics and mergers. 
Although these have been identified in previous works \citep[e.g.][]{2021Vulcani,2025Poggianti}, we do not exclude them a priori, both to test whether a rotating disc reproduces at least part of their kinematics and to maximise the sample size. 
Most previously identified mergers are eventually excluded, because neither the gas nor the stellar kinematics are consistent with a rotating disc, or because of low $i$, low S/N, or $V_\mathrm{LOS}$ below $\sigma_\mathrm{MUSE}$; the reason is given galaxy by galaxy in Table~\ref{tab:sample}, and discussed in the text when necessary. 
We stress that our selection criteria are meant to maximise the final sample size, optimising the balance between unperturbed and perturbed galaxies with sufficiently robust kinematics. 
This pushes the tilted-ring model to its limits, but preserves as much as possible the signal of environmental effects. 

\subsection{Stellar kinematics}\label{sec:StarKin}
We measure the stellar rotation velocity $V_\mathrm{rot,\star}$ with the \barolo task \texttt{2DFIT}, which fits the velocity field pixel-by-pixel with a tilted-ring model, where $	V_\mathrm{LOS,\star} = V_\mathrm{sys} + V_\mathrm{rot,\star} \sin i \cos \phi$ assuming $V_\mathrm{rad,\star} = 0$~\kms. 
For each galaxy, we perform a sequence of three fits in which the geometry and $V_\mathrm{sys}$ are progressively fixed, so that $V_\mathrm{rot,\star}$ is the only free parameter in the final model. 
In all runs we set \texttt{radsep=0.8}, spacing rings by 0.8\arcsec, about 20\% below the MUSE seeing: adjacent rings are then not fully independent, especially where the Voronoi bins are larger, but velocity gradients are well sampled. 
The number of rings is set, for each galaxy, to cover the largest area with a clear velocity gradient and good S/N.
The uncertainty on $V_\mathrm{rot,\star}$ combines that on $i$ and the statistical scatter of the residuals (details in Appendix~\ref{app:errors}). 
For each galaxy, we inspect the residuals between best-fit model and observed velocity field (panels I, J, and K in \citetalias{myGASPOnlineAtlas}). 
The data are well reproduced, with residuals below $\sigma_\mathrm{MUSE}$ and no systematic pattern. 
Only some galaxies with prominent bars (e.g. JO194) show central patterns typical of non-circular motions, which our model based on circular orbits cannot capture. 
In Appendix~\ref{app:kin_setup}, we perform a harmonic analysis to confirm that our results are not influenced by the adopted modelling procedure. 
We exclude from further analysis galaxies with large residuals indicating that the kinematics is inconsistent with a rotating disc (e.g. JO13), mostly confirmed mergers \citep[see][]{2021Vulcani,2025Poggianti} and galaxies with low S/N (see Table~\ref{tab:sample}). 

Radial profiles of the stellar velocity dispersion $\sigma_\star(R)$ are obtained by dividing velocity dispersion maps into annuli with the photometric geometry and computing $\sigma_\star$ and its uncertainty as the median and $1.48 \times \mathrm{MAD}$ of the Voronoi bins with centroid inside the ring.
Profiles and maps are shown in panels L and O (see \citetalias{myGASPOnlineAtlas}). 
We measure $V_\mathrm{rot,\star}$ and $\sigma_\star$ for 82 galaxies in total, 49 in clusters and 33 in groups or the field. 
These 2D approaches do not account for beam smearing, which inflates $\sigma_\star$ and smooths $V_\mathrm{rot,\star}$ gradients (Appendix~\ref{app:errors}). 
This bias worsens for high $i$, steep $V_\mathrm{LOS}$ gradients, and poor resolution, possibly introducing a dependence on $M_\star$ through the mass-size relation. 
Most of our galaxies, however, are resolved with $>10$ resolution elements along the major axis and, beyond one-two resolution elements, $V_\mathrm{circ}$ is well recovered (Appendix~\ref{app:kin_setup}), also because the underestimate of $V_\mathrm{rot}$ is partly compensated by the overestimate of $\sigma_\star$ and thus $V_\mathrm{AD,\star}$.
Beam smearing is anyway independent of environment, so its impact on our conclusions is at most secondary. 

\subsection{Ionised gas kinematics}\label{sec:GasKin}
We measure the ionised gas kinematics with the \barolo task \texttt{3DFIT}, which fits the tilted-ring model to emission line datacubes channel by channel. 
Prior to the fitting, the model is convolved with the telescope point spread function (PSF) and the line spread function (LSF; assumed Gaussian), in order to account for beam smearing and instrumental broadening. 
We assume a circular Gaussian PSF with FWHM~$=1$\arcsec, the MUSE seeing. 
Appendix~\ref{app:kin_setup} details our \texttt{3DFIT} setup and discusses the consequences of the assumed PSF and LSF shapes. 
\barolo\ also masks the datacube to keep only the galaxy emission: although dispensable at our S/N, we apply the masking to reduce the number of fitted spaxels, and hence the computational time, and to remove \nii\ emission in a few galaxies. 
We run \texttt{3DFIT} on the \halpha\ and \hbeta\ datacubes. 
With the latter, we assess whether \nii\ contamination would affect the best-fit model, which is not the case for most galaxies, since \nii\ is much fainter than \halpha. 

The initial guesses for $x_0$, $y_0$, $\phi$, and $i$ come from the isophote fitting, with \texttt{radsep=0.8}\arcsec. 
The number of rings covers the largest axisymmetric area, without pushing the fit to radii where few spaxels contain flux. 
Regions contaminated by the AGN-driven outflows are excluded after inspecting the PVDs. 
The free parameters are $V_\mathrm{rot,\hii}$ and $\sigma_\mathrm{\hii}$, and $\phi$ and/or $i$ (to account for warps).
We let \barolo compute $V_\mathrm{sys}$ automatically, adjusting it when needed to make the PVD quadrants as symmetric as possible. 
Most galaxies are axisymmetric enough to fit both sides at once. 
Some, however, show bright star-forming clumps where lines are broader and the kinematics seems "incoherent" with the rest of the disc; in these cases, we model only the side with fewer clumps. 
This choice does not bias our results (see Sect.~\ref{sec:AMR_env}), since it applies to only nine galaxies in the fiducial sample (not preferentially for cluster members) and the excluded side is chosen because of clumps in the inner disc, not at its edge where environmental effects are expected. 
Indeed, the PVDs show that the model reproduces well the unfitted side too. 
In our set up, the fitting gives more weight to the pixels near the major axis and minimises the absolute residuals between data and model, normalised to the azimuthally-averaged flux of each ring, further reducing the impact of clumps and other bright structures. 

By inspecting channel maps and PVDs (panels R and V in \citetalias{myGASPOnlineAtlas}), we verify that the best-fit models reproduce the observations for most galaxies. 
The exceptions, excluded from further analysis (Table~\ref{tab:sample}), have signatures of a major merger or are in very advanced stages of gas removal (e.g. JW56). 
The \halpha\ emission is nonetheless reproduced reasonably well for P3984, a faint low mass galaxy classified as a merger by \cite{2021Vulcani}, and for the merger candidate P877 \citep{2021Vulcani}, whose kinematics is however uncertain because $i \approx 32\deg$. 
We can usually model at least one side of the ionised disc even for strongly stripped and jellyfish galaxies, which therefore remain rotation-dominated. 
We check that the best-fit $V_\mathrm{rot, \hii}$ and $\sigma_\mathrm{\hii}$ are global minima using the \barolo task \texttt{spacepar} for a few galaxies with large $V_\mathrm{LOS,\hii}$ and for the most problematic ones, which are those with low mass and $i$, since $V_\mathrm{LOS,\hii}$ may fall below $\sigma_\mathrm{MUSE}$. $V_\mathrm{rot,\hii}$ is generally well constrained at all $R$. 
Exceptions are JO119b, P17048b, JO179, and P63947 (\citealt{2021Vulcani} classify the last two as mergers), which are excluded from further analysis: since $V_\mathrm{LOS,\hii}$ is low, it covers too few channels to measure the kinematics reliably. 
Overall, 79 galaxies have reasonably robust $V_\mathrm{rot,\hii}(R)$, 48 in clusters and 31 in the field.

\subsection{Asymmetric drift correction and circular velocity curves}\label{sec:VAD_Vcirc}
Many works derive the TFRs from \hi\ rotation curves, where the pressure support against gravity is negligible and $V_\mathrm{rot,\hi}(R)$ traces the circular velocity $V_\mathrm{circ}(R)$ well, at least for sufficiently massive galaxies \citep[e.g.][]{2017Iorio}. 
However, it is not so for our tracers, especially stars \citep[e.g.][]{2013Martinsson}, for which $V_\mathrm{circ}$ must be computed by correcting $V_\mathrm{rot}$ for the asymmetric drift velocity $V_\mathrm{AD}$ \citep{2008BinneyTremaine}:
\begin{equation}\label{eq:vc_vad}
	V_\mathrm{circ}^2(R) = V_\mathrm{rot}^2(R) + V_\mathrm{AD}^2(R) \,.
\end{equation}
Under the simplifying assumption that the disc scale height is constant with radius, we have
\begin{equation}\label{eq:vad}
	V_\mathrm{AD}^2(R) = - \sigma^2(R) \frac{\partial \ln \left( \Sigma(R) \sigma^2(R) \right) }{\partial \ln R} \,.
\end{equation}
When computing Eq.~\ref{eq:vad}, we regularise $\Sigma(R)$ and $\sigma(R)$ with models reproducing the observed radial decline, to avoid local variations (e.g. peaks from star-forming clumps or drops in low S/N regions) that would produce spurious discontinuities in $V_\mathrm{circ}(R)$. 
We model $\Sigma_\mathrm{gas}$ and $\sigma_\mathrm{\hii}$ as exponentials described by the central value and a scale radius. 
We use the model of Sect.~\ref{sec:StarDist} for $\Sigma_\star$ and an exponential for $\sigma_\star$, adding where necessary a floor dispersion to reproduce the flattening of $\sigma_\star(R)$ at large $R$. 
In the \citetalias{myGASPOnlineAtlas}, the best-fit models for $\Sigma_\star(R)$, $\sigma_\star(R)$, and $\sigma_\mathrm{\hii}(R)$ are shown in panels D, O, and W. 
$V_\mathrm{AD}$, $V_\mathrm{circ}$ and their uncertainties are computed with a Monte Carlo approach, propagating the uncertainties on $\Sigma(R)$ and $\sigma(R)$ models, and on $V_\mathrm{rot}$ (Appendix~\ref{app:errors}). 

Equation~\ref{eq:vad} implicitly assumes that both $\sigma_\star$ and $\sigma_\mathrm{\hii}$ are isotropic. 
This approximation is reasonable for the ionised gas \citep{2022Law}, but stellar velocities are known to be anisotropic \citep[e.g.][]{2008Noordermeer,2017Marchuk,2018Chemin,2018Pinna,2019MogotsiRomeo}. 
The observed $\sigma_\star$ results from the combination of projection effects along the LOS, anisotropy, and the tilt of the velocity dispersion tensor, which are not known a priori and may vary within and between galaxies. 
We therefore assume isotropy, which seems reasonable since our $V_\mathrm{circ}(R)$ from $V_\mathrm{rot,\star}(R)$ agrees, within uncertainties and local fluctuations, with that from $V_\mathrm{rot,\hii}(R)$. 

For star forming galaxies, we adopt ionised gas or stars as fiducial kinematic tracer whichever reproduces the observations out to the largest radius, based on the inspection of the best-fit models of the stellar velocity field and of the gas datacube.
We choose \halpha\ for most galaxies, since it is brighter than \hbeta\ and stellar continuum. 
We also prefer ionised gas over stars as kinematic tracers because $V_\mathrm{rot,\hii}$ is corrected for beam smearing effect while $V_\mathrm{rot,\star}$ is not. 
Where \nii\ contamination is significant, we adopt \hbeta\ or the stars, whichever rotation curve reaches larger $R$. 
For passive discs (S0 and S0/a), which lack an extended \halpha\ disc, the stars are the only tracer available. 
In a few cases, we combine tracers preferring, for instance, $V_\mathrm{rot,\star}$ where the central regions are contaminated by outflows and winds. 
These fiducial rotation curves are then used to compute $V_\mathrm{circ}$ via Eq.~\ref{eq:vc_vad}. 

\section{Results}\label{sec:results}
We now present our fiducial sample with reliable kinematics (Sect.~\ref{sec:fiducial_sample}), analysed in Sect.~\ref{sec:bar_kin_env}, and derive the TFRs (Sect.~\ref{sec:TFR}) and SAMRs (Sect.~\ref{sec:AMR}). 
In logarithmic space, these scaling relations are power laws with form $\log y = \alpha \log x +\beta$, where $\alpha$ and $\beta$ are the slope and $y$-intercept. 
Appendix~\ref{app:orto_lin_reg} details the orthogonal linear regression, which fits for two parameters ($\theta$ and $\gamma$) related to $\alpha$ and $\beta$, and the orthogonal intrinsic scatter of the relation $\delta_\perp$, used to compute the vertical scatter $\delta_y$ (see Table~\ref{tab:scaling_bestfit}). 
\begin{figure*}[!t]
	\centering
	\includegraphics[width=1\textwidth]{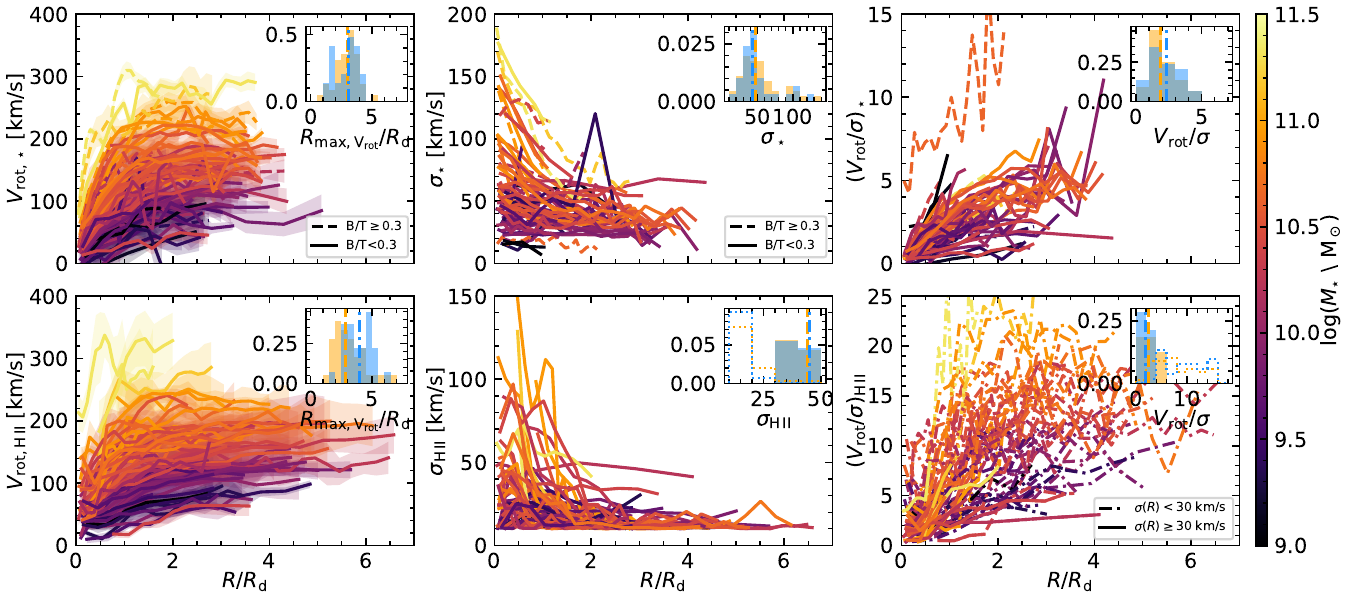}
	\caption{Radial profiles of rotation velocity (left), velocity dispersion (centre), and their ratio (right) for stars (top) and ionised gas (bottom), coloured by stellar mass. 
	Dashed curves (top left panel) are galaxies with a significant bulge; dotted curves (bottom right panel) are radii with $\sigma < 30$~\kms. 
	The insets show the distributions of the maximum radius with measured $V_\mathrm{rot}$ ($R_\mathrm{max,V_\mathrm{rot}}$), in units of disc scale length $R_\mathrm{d}$, and of the average $\sigma$ and $V_\mathrm{rot}/\sigma$, for cluster (yellow) and field (blue) galaxies; filled histograms use only $\sigma \geq 30$~\kms, empty ones all measurements. 
	For readability, uncertainties (shaded bands) are shown only for $V_\mathrm{rot}$. }
	\label{fig:V_over_sigma}
\end{figure*}

\subsection{The fiducial sample}\label{sec:fiducial_sample}
Hereafter, we consider only discs with $30 \lesssim i/\deg \lesssim 75$\footnote{We also remove JO149, which looks face on but is too faint and irregular to robustly derive its inclination.}. 
At lower $i$, the velocity component along the LOS is small and errors on $i$ propagate strongly onto $V_\mathrm{rot}$. 
For nearly edge on discs, instead, the tilted-ring modelling becomes less reliable, because regions with different kinematics overlap along the LOS \citep[][]{2015DiTeodoro}. 
The same overlap of regions with different extinction makes also the Balmer decrement uncertain and prone to self-obscuration \citep[e.g.][]{2019Concas}; selecting against high $i$ avoids systematics due to large extinction corrections. 
This inclination cut also mitigates the selection bias of the parent sample (Sect.~\ref{sec:class}), making the median $i$ of the stripped galaxies similar to that of the undisturbed ones. 
A threshold of $70\deg$ would remove many stripped galaxies without reducing this difference further. 
Galaxies excluded by this inclination cut are reported in Table~\ref{tab:sample}. 

Most star-forming galaxies have both gas and stellar kinematics; the eight with gas only are used for the TFRs and the gas SAMR but not for the baryonic SAMR, where $j_\mathrm{bar}$ is dominated by $j_\star$ (see Table~\ref{tab:sample}). 
For the jellyfish galaxy JW56, whose gas kinematics is not robust, and for the passive discs with no \halpha, we adopt $V_\mathrm{circ,\star}$ for the TFRs and include them only in the stellar and baryonic SAMR, implicitly assuming no cold gas is left. 
The datasets for the TFRs and SAMRs are therefore slightly different, without consequences for our conclusions. 
The TFRs are derived for 81 galaxies, the stellar SAMR for 73, the gas SAMR for 49, and the baryonic SAMR for 71. 
Of the 37 galaxies excluded from the parent sample, 24 have unfavourable inclination (12 too face-on, 8 too edge-on, 4 with low $i$ plus low S/N or $V_\mathrm{LOS}$ below resolution) and the remaining 13 are mostly mergers. 
The selection is therefore primarily geometrical, and does not rely on the physical properties that could carry the environmental signal. 
The fiducial sample remains representative of the GASP one: our galaxies lie within the scatter of the mass--size relation of \citet{2020Franchetto} and, using the measurements of \citet{2018Vulcani_b}, we recover a mean SFR enhancement in stripped galaxies of $\approx 0.15$~dex compared to the SFR--$M_\star$ relation of the unperturbed ones, compatible with that found for the full GASP sample. 
We caution that scaling relations are sensitive to sample selection and completeness: GASP is a targeted survey, so our conclusions concern the relative position of GASP and its subsamples rather than the absolute slopes and normalisations.

\subsection{Rotation and velocity dispersion profiles}\label{sec:bar_kin_env}
The top panels of Fig.~\ref{fig:V_over_sigma} show $V_\mathrm{rot,\star}(R)$, $\sigma_\star(R)$, and their ratio $(V_\mathrm{rot}/ \sigma)_\star$. 
As expected, $V_\mathrm{rot,\star}$ increases with $M_\star$, reaching $\approx 250-300$~\kms\ in the most massive galaxies. 
For $84$\% of the sample, the maximum radius with measured $V_\mathrm{rot}$ ($R_\mathrm{max,V_\mathrm{rot}}$) lies between $\approx 1.9$ and $\approx 3.8$ disc scale lengths ($R_\mathrm{d}$). 
Also $\sigma_\star$ increases with $M_\star$, declining nearly exponentially with $R$ and flattening at large radii. 
The ratio $(V_\mathrm{rot}/ \sigma)_\star$ is of order unity in the inner disc, so pressure support is not negligible for the stars. 
The insets show that the distributions of $R_\mathrm{max,V_\mathrm{rot}}/R_\mathrm{d}$, of the average $\sigma_\star$, and of $(V_\mathrm{rot}/ \sigma)_\star$ are indistinguishable for cluster and field galaxies, with medians over the whole sample $R_\mathrm{max,V_\mathrm{rot}}/R_\mathrm{d} = 3.0_{-1.1}^{+0.7}$, $\sigma_\star = 44_{-12}^{+30}$~\kms, and $(V_\mathrm{rot}/ \sigma)_\star = 2.0_{-0.9}^{+1.5}$. 
The environmental effect on stellar kinematics, if any, is therefore minor. 

The bottom panels show the ionised gas, with very similar trends. 
For $84$\% of the sample the maximum radius with measured $V_\mathrm{rot, \hii}$ lies between $\approx 2.4$ and $\approx 4.9 R_\mathrm{d}$, so the gas rotation curves are slightly more extended than the stellar ones. 
The inset shows that cluster and field galaxies have, respectively, the median $R_\mathrm{max,V_\mathrm{rot}}/R_\mathrm{d} = 2.8_{-0.8}^{+1.6}$ and $R_\mathrm{max,V_\mathrm{rot}}/R_\mathrm{d} = 4.0_{-1.2}^{+0.9}$. 
The distributions for cluster and field differ: a Kolmogorov-Smirnov (KS) test gives a probability of no correlation of $p_\mathrm{null} \approx 0.03$ and a Mann-Whitney test $p_\mathrm{null} \approx 0.007$. 
Instead, the distributions for stars are statistically indistinguishable ($p_\mathrm{null} = 0.90$), indicating that the
effect is specific to the cluster environment and the gas only, with important consequences for $j_\mathrm{gas}$ (Sec.~\ref{sec:AMR}). 
Differences between $V_\mathrm{rot,\star}$ and $V_\mathrm{rot, \hii}$ are typically explained by the different level of pressure support for the two components (see Sect.~\ref{sec:VAD_Vcirc}) and do not correlate with the degree of ram pressure stripping. 
An instructive exception is the jellyfish galaxy JO206 (see also JO162; \citetalias{myGASPOnlineAtlas}) where, beyond the bar extent ($R \gtrsim 6-7$~kpc), $V_\mathrm{rot,\hii}$ is lower than $V_\mathrm{rot,\star}$ even though the ionised gas disc preserves some coherent kinematics. 

The bottom central panel shows $\sigma_\mathrm{\hii}(R)$, which spans the range $20\text{\kms} \lesssim \sigma_\mathrm{\hii} \lesssim 50\text{\kms}$ (slightly lower than $\sigma_\star$) and correlates with $M_\star$ \citep[see also][]{2020Varidel,2022Law} similarly to but much more weakly than $\sigma_\star(R)$. 
As found for cold gas tracers \citep[e.g.][]{2017Iorio,2019Bacchini,2020Bacchini}, $\sigma_\mathrm{\hii}$ decreases with increasing $R$. 
The parameter space inspection (see Appendix~\ref{app:kin_setup}) reveals that $\sigma_\mathrm{\hii}$ is often an upper limit when $\lesssim 30-40$~\kms: with our methodology, the MUSE spectral resolution is insufficient to measure it, except where $\sigma_\mathrm{\hii} \gtrsim 30$~\kms\ and the S/N is high, typically $R \lesssim 2 R_\mathrm{d}$. 
The inset shows that, excluding the rings with $\sigma_\mathrm{\hii} < 30$~\kms, the disc-averaged $\sigma_\mathrm{\hii}$ of cluster and field galaxies remain very similar, with median $\approx 44-45$~\kms. 
By inspecting the PVDs and $\sigma_\mathrm{\hii}(R)$, we find no obvious local dependence on the degree of stripping either, not even under strong or extreme stripping.

The bottom right panel shows that $(V_\mathrm{rot}/ \sigma)_\mathrm{\hii}$ increases with $M_\star$ like $(V_\mathrm{rot}/ \sigma)_\star$ but reaches higher values, so the asymmetric drift correction matters less for the gas ($ V_\mathrm{AD,\hii}^2 / V_\mathrm{rot,\hii}^2 \lesssim 0.1$ even at $\log M_\star/\mathrm{M}_\odot \lesssim 9.5$) than for the stars, for which it is $\approx3$ times larger. 
Most measurements at $ (V_\mathrm{rot}/ \sigma)_\mathrm{\hii} \gtrsim 5$ come from regions with $\sigma_\mathrm{\hii} < 30$~\kms\ and are therefore lower limits. 
The cluster and field distributions of $(V_\mathrm{rot}/ \sigma)_\mathrm{\hii}$ differ only marginally, the high tail of the clusters giving a slightly larger average of $\approx 2.4$ against $\approx 1.9$. 
The environmental effect on the global kinematics of the ionised disc is thus negligible, or too small to detect with our data and methods within the area we model. 
Appendix~\ref{app:Vsigma_rps} shows that $(V_\mathrm{rot}/ \sigma)_\mathrm{\hii}$ mildly decreases with the degree of stripping, but the correlation is not statistically significant. 
We do not detect the local $\sigma_\mathrm{\hii}$ enhancements reported by some studies \citep[e.g.][]{2026Brown}, perhaps because the tilted-ring modelling is not optimal to detect them. 
Such a task requires a pixel-by-pixel modelling of the line profiles to separate turbulence from anomalous motions \citep[e.g.][]{2023Marasco} and higher spatial resolution to reduce beam smearing. 
Our models nonetheless reproduce well the observations, even for strong and extreme stripping cases (e.g. A3266\_B\_0257, JO85), which disfavours a systematic underestimation of $\sigma_\mathrm{\hii}$.

\subsection{Tully-Fisher relations}\label{sec:TFR}
The TFR reads
\begin{equation}\label{eq:TFR}
	\log \left( \frac{M}{\mathrm{M_\odot}}\right) = \alpha \log \left(\frac{V_\mathrm{flat}}{\mathrm{km~s}^{-1}}\right) + \beta \,,
\end{equation}
where $M$ is $M_\star$ and $M_\mathrm{bar}$ for the stellar and baryonic TFRs, respectively, and $V_\mathrm{flat}$ is the velocity along the flat part of $V_\mathrm{circ}(R)$. 
This definition minimises the baryonic TFR scatter \citep{2001Verheijen,2019Lelli}. 
We compute $V_\mathrm{flat}$ with the iterative procedure of \cite{2016Lelli}: the mean velocity $V_\mathrm{ave}$ is computed from the two outermost points of $V_\mathrm{circ}(R)$ and then proceeding inward, adding more points until two subsequent calculations differ less than 5\% (see Appendix~\ref{app:errors} for uncertainties). 
Where this criterion is not met, we use the outermost points of $V_\mathrm{circ}(R)$, reasonably close to $V_\mathrm{flat}$ for galaxies with $M_\mathrm{bar} \lesssim 10^{10}$~\msun\, and rising $V_\mathrm{circ}(R)$ (but see Sect.~\ref{sec:literature_TFR}). 
This occurs for 20 galaxies ($\approx 25$\% of the sample) but we have verified it does not impact our conclusions. 
Eventually, 81 galaxies (51 in clusters and 30 in low density environments) have $V_\mathrm{flat}$ measurements from either the stellar or the ionised gas kinematics. 

\begin{table}
	\centering
	\caption{Best-fit scaling relations for the fiducial sample (all), cluster and field galaxies, and ram pressure stripped and unperturbed (control) galaxies. $\alpha$ and $\beta$ are the slope and intercept, $\delta_\perp$ and $\delta_y$ the orthogonal and vertical intrinsic scatter.}
	\label{tab:scaling_bestfit}
	\begin{tabular}{c c c c c}
		\hline\hline
		Sample & $\alpha$ & $\beta$ & $\delta_\perp$ & $\delta_y$ \\ \hline\hline
		\multicolumn{5}{c}{Stellar TFR} \\ \hline
		All & $3.58_{-0.21}^{+0.22}$ & $2.43_{-0.47}^{+0.45}$ & $0.06_{-0.01}^{+0.01}$ & $0.21_{-0.03}^{+0.04}$ \\
		Cluster & $3.55_{-0.23}^{+0.28}$ & $2.48_{-0.60}^{+0.52}$ & $0.06_{-0.01}^{+0.01}$ & $0.21_{-0.04}^{+0.05}$ \\
		Field & $3.57_{-0.38}^{+0.50}$ & $2.46_{-1.06}^{+0.86}$ & $0.06_{-0.01}^{+0.02}$ & $0.22_{-0.05}^{+0.07}$ \\
		Stripped & $3.66_{-0.22}^{+0.27}$ & $2.23_{-0.61}^{+0.47}$ & $0.05_{-0.01}^{+0.01}$ & $0.19_{-0.03}^{+0.04}$ \\
		Control & $3.30_{-0.36}^{+0.48}$ & $3.11_{-1.02}^{+0.78}$ & $0.07_{-0.01}^{+0.02}$ & $0.23_{-0.05}^{+0.07}$ \\
		\hline
		\multicolumn{5}{c}{Baryonic TFR} \\ \hline
		All & $3.48_{-0.20}^{+0.23}$ & $2.74_{-0.49}^{+0.43}$ & $0.06_{-0.01}^{+0.01}$ & $0.22_{-0.03}^{+0.04}$ \\
		Cluster & $3.49_{-0.23}^{+0.29}$ & $2.71_{-0.65}^{+0.51}$ & $0.06_{-0.01}^{+0.01}$ & $0.21_{-0.04}^{+0.05}$ \\
		Field & $3.37_{-0.37}^{+0.52}$ & $3.01_{-1.13}^{+0.80}$ & $0.06_{-0.01}^{+0.02}$ & $0.23_{-0.05}^{+0.07}$ \\
		Stripped & $3.60_{-0.21}^{+0.26}$ & $2.45_{-0.57}^{+0.45}$ & $0.05_{-0.01}^{+0.01}$ & $0.19_{-0.03}^{+0.04}$ \\
		Control & $3.10_{-0.36}^{+0.47}$ & $3.61_{-1.01}^{+0.76}$ & $0.07_{-0.01}^{+0.01}$ & $0.22_{-0.05}^{+0.07}$ \\
		\hline
		\multicolumn{5}{c}{Stellar SAMR} \\ \hline
		All & $0.68_{-0.05}^{+0.05}$ & $-4.12_{-0.51}^{+0.52}$ & $0.15_{-0.01}^{+0.02}$ & $0.19_{-0.02}^{+0.02}$ \\
		Cluster & $0.70_{-0.05}^{+0.06}$ & $-4.26_{-0.58}^{+0.52}$ & $0.13_{-0.01}^{+0.02}$ & $0.16_{-0.02}^{+0.02}$ \\
		Field & $0.67_{-0.10}^{+0.11}$ & $-4.01_{-1.06}^{+1.01}$ & $0.18_{-0.02}^{+0.03}$ & $0.22_{-0.03}^{+0.04}$ \\
		Stripped & $0.67_{-0.05}^{+0.05}$ & $-3.97_{-0.51}^{+0.50}$ & $0.14_{-0.01}^{+0.02}$ & $0.16_{-0.02}^{+0.02}$ \\
		Control & $0.77_{-0.12}^{+0.14}$ & $-5.10_{-1.41}^{+1.23}$ & $0.18_{-0.02}^{+0.03}$ & $0.23_{-0.03}^{+0.05}$ \\
		\hline
		\multicolumn{5}{c}{Gas SAMR} \\ \hline
		All & $0.77_{-0.10}^{+0.10}$ & $-4.49_{-0.92}^{+0.91}$ & $0.16_{-0.02}^{+0.02}$ & $0.20_{-0.02}^{+0.03}$ \\
		Cluster & $0.71_{-0.10}^{+0.11}$ & $-4.05_{-1.01}^{+0.91}$ & $0.15_{-0.02}^{+0.03}$ & $0.18_{-0.03}^{+0.03}$ \\
		Field & $0.80_{-0.17}^{+0.20}$ & $-4.69_{-1.85}^{+1.61}$ & $0.15_{-0.03}^{+0.04}$ & $0.19_{-0.04}^{+0.06}$ \\
		Stripped & $0.71_{-0.10}^{+0.12}$ & $-4.05_{-1.10}^{+0.97}$ & $0.15_{-0.02}^{+0.03}$ & $0.18_{-0.03}^{+0.04}$ \\
		Control & $0.80_{-0.14}^{+0.15}$ & $-4.74_{-1.45}^{+1.32}$ & $0.14_{-0.03}^{+0.03}$ & $0.18_{-0.04}^{+0.05}$ \\
		\hline
		\multicolumn{5}{c}{Baryonic SAMR} \\ \hline
		All & $0.68_{-0.04}^{+0.05}$ & $-4.19_{-0.51}^{+0.44}$ & $0.12_{-0.01}^{+0.01}$ & $0.15_{-0.02}^{+0.02}$ \\
		Cluster & $0.69_{-0.05}^{+0.05}$ & $-4.27_{-0.51}^{+0.46}$ & $0.09_{-0.02}^{+0.02}$ & $0.11_{-0.02}^{+0.02}$ \\
		Field & $0.65_{-0.09}^{+0.10}$ & $-3.90_{-1.00}^{+0.89}$ & $0.15_{-0.02}^{+0.03}$ & $0.18_{-0.03}^{+0.04}$ \\
		Stripped & $0.67_{-0.04}^{+0.04}$ & $-3.97_{-0.45}^{+0.43}$ & $0.10_{-0.01}^{+0.02}$ & $0.12_{-0.02}^{+0.02}$ \\
		Control & $0.75_{-0.11}^{+0.12}$ & $-4.87_{-1.25}^{+1.12}$ & $0.15_{-0.02}^{+0.03}$ & $0.18_{-0.03}^{+0.04}$ \\
		\hline\hline
		\end{tabular}
\end{table}

\subsubsection{Environmental dependence of the TFRs}\label{sec:TFR_env}
\begin{figure*}
	\centering
	\includegraphics[width=1\textwidth]{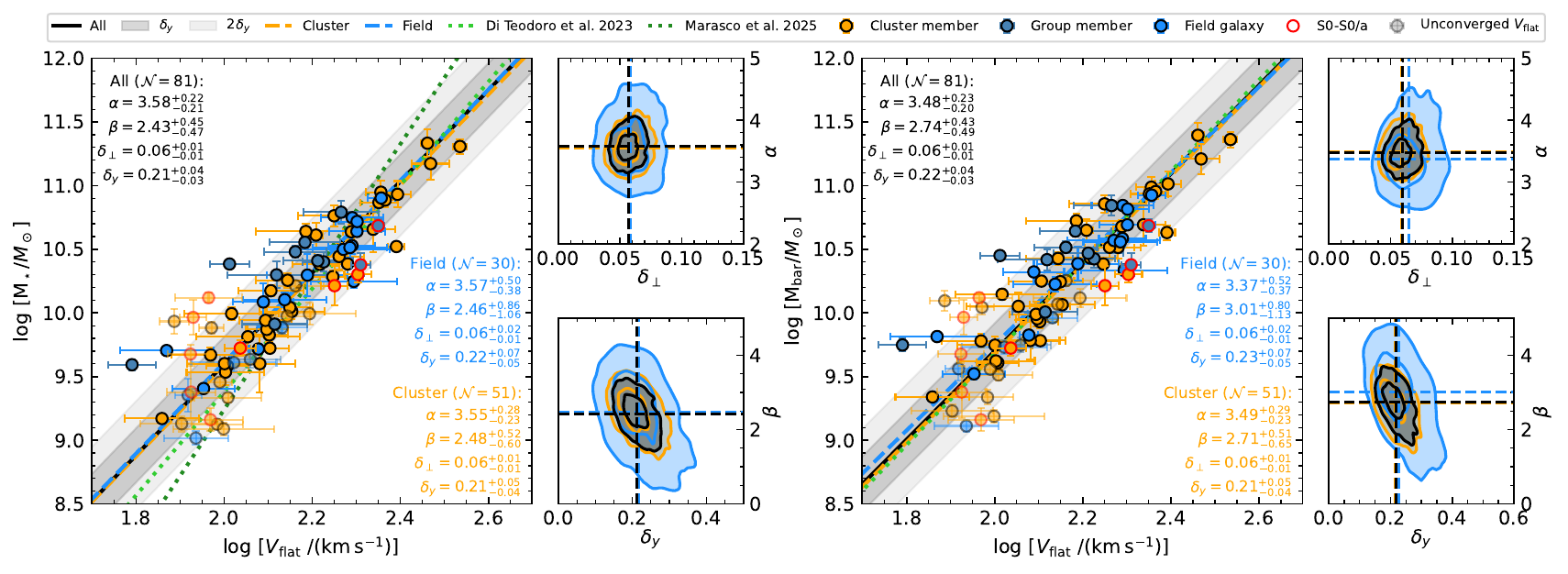}
	\caption{Stellar (left) and baryonic (right) TFRs, with 2D posteriors of the best-fit parameters (small panels). 
		Orange and blue colours are for the cluster and field sub-samples, respectively; shaded symbols mark galaxies with non-converged $V_\mathrm{flat}$ but included in the fitting. 
		Black lines and grey bands are the best fit and scatter of the fiducial sample, dashed lines the fits of the two sub-samples. 
		Text boxes report the parameters, with $\mathcal{N}$ the number of galaxies. 
		Green lines are $V_\mathrm{flat,\hi}$-based fits from the literature (Sect.~\ref{sec:literature_TFR}). 
		In the posterior panels, dashed lines mark the median, contours enclose $\delta$ and $2\delta$ of the probability. 
		The cluster and field TFRs are essentially indistinguishable, indicating no environmental dependence.}
	\label{fig:TFR}
\end{figure*}

Figure~\ref{fig:TFR} shows the stellar and baryonic TFRs for the fiducial sample, cluster members, and field and group galaxies. 
The stellar TFR of the fiducial sample has $\alpha \approx 3.58$, $\beta \approx 2.43$, $\delta_\perp \approx 0.06$ and $\delta_y \approx 0.21$. 
The sub-samples give very similar values, $\alpha \approx 3.55$ for cluster members and $\approx 3.57$ for field galaxies with essentially the same scatter; indeed their 2D posteriors in Fig.~\ref{fig:TFR} overlap. 
Galaxies just outside $\pm 2 \delta_\perp$ are not genuine outliers but problematic cases with uncertain and/or low $i$, potential minor merger, or relatively low S/N. 
As expected, the baryonic TFR is shallower than the stellar TFR, with $\alpha \approx 3.48-3.49$ for the whole sample and cluster members, $\alpha \approx 3.37$ for field galaxies, and $\delta_\perp \approx 0.06$ and $\delta_y \approx 0.21 - 0.23$.
All differences between the three best-fits are $< 1\delta$. 
Group members follow the same trends without a preferential location in both TFRs, and Appendix~\ref{sec:TFR_rps} shows that the results hold when comparing stripped against control galaxies and when using different kinematic tracers. 
The best-fit parameters alone, however, do not quantify how large an environmental effect must be to be detectable. 
Hence, in Appendix~\ref{app:orto_lin_reg}, we also measure the mean vertical offset between sub-samples $\Delta_y$ and its uncertainty $\delta_\Delta$, that is the typical offset that would arise by chance alone given our sample size and scatter. 
Over the common $V_\mathrm{flat}$ range, cluster and field galaxies are offset by $\Delta_y = -0.02\pm0.06$~dex for both the stellar and the baryonic TFR, and stripped against control galaxies by $-0.06\pm0.07$~dex: all consistent with zero and lower than $\delta_y$.
We conclude that the environment imprints are negligible or undetectable with our sample and methodologies.

\subsection{Specific angular momentum - mass relations}\label{sec:AMR}
For a rotating disc, the definition of $j$ is
\begin{equation}\label{eq:j_1comp}
	j(<R) = \frac{J}{M} = \frac{2 \pi \int_{0}^{R} \Sigma(R') V_\mathrm{rot}(R') R'^2 dR'}{2 \pi \int_{0}^{R} \Sigma(R') R' dR'} \,.
\end{equation}
We calculate $j_\star$ using the exponential disc model of $\Sigma_\star$ (Sect.~\ref{sec:StarDist}) and $V_\mathrm{rot,\star}(R)$. 
Since the $M$--$j$ correlation is tighter for the disc alone \citep{2018Posti}, we exclude bulges, which add mass at small $R$ and lower $j_\star$; since most galaxies have $B/T \lesssim 0.4$, including bulges leaves our main conclusions unchanged (see Appendix~\ref{app:j_bulge}). 
We extrapolate the exponential disc model to $R_\mathrm{max}$ from photometry, to cover as much as possible of the outer disc that carries a large fraction of $j_\star$ \citep{2012RomanowskyFall}. 
Where $V_\mathrm{rot,\star}$ is not measured directly, we assume it constant at the flat/outermost value using the criterion for $V_\mathrm{flat}$ (see Sect.~\ref{sec:TFR}). 
We compute $j_\mathrm{gas}$ from $\Sigma_\mathrm{gas}$ and $V_\mathrm{circ,\hii}$, without extrapolating beyond the outermost radius where $\Sigma_\mathrm{gas}$ is measured, to avoid considering regions with no gas, because either not detected in \halpha\ or stripped. 
Our $j_\mathrm{gas}$ should thus approximate the cold gas $j$, for which pressure support is negligible and $V_\mathrm{rot,gas} \approx V_\mathrm{circ}$. 
For galaxies with measured $j_\mathrm{\star}$
\footnote{Stars and gas counter-rotate in P4946 \citep{2018Vulcani}, so $J_\star$ has opposite sign to $J_\mathrm{gas}$ in Eq.~\ref{eq:jbar}. 
The low $S/N$ allows measuring the stellar kinematics only for $R < 3$~kpc, so we cannot tell whether the inner counter-rotating component is surrounded by a co-rotating disc.}, $j_\mathrm{bar}$ is
\begin{equation}\label{eq:jbar}
	j_\mathrm{bar}(<R) = \frac{J_\mathrm{bar}}{M_\mathrm{bar}} = \frac{J_\star + J_\mathrm{gas}}{M_\mathrm{bar}}\,.
\end{equation}
We use a Monte Carlo approach to compute $\delta j$ (see Appendix~\ref{app:errors}). 
When fitting the SAMRs, we include only galaxies whose $j$ calculation has converged according to the criterion defined by \cite{2021ManceraPina} (see Appendix~\ref{app:j_convergence}). 

\subsubsection{Environmental dependence of the SAMRs}\label{sec:AMR_env}
\begin{figure*}
	\centering
	\includegraphics[width=1\textwidth]{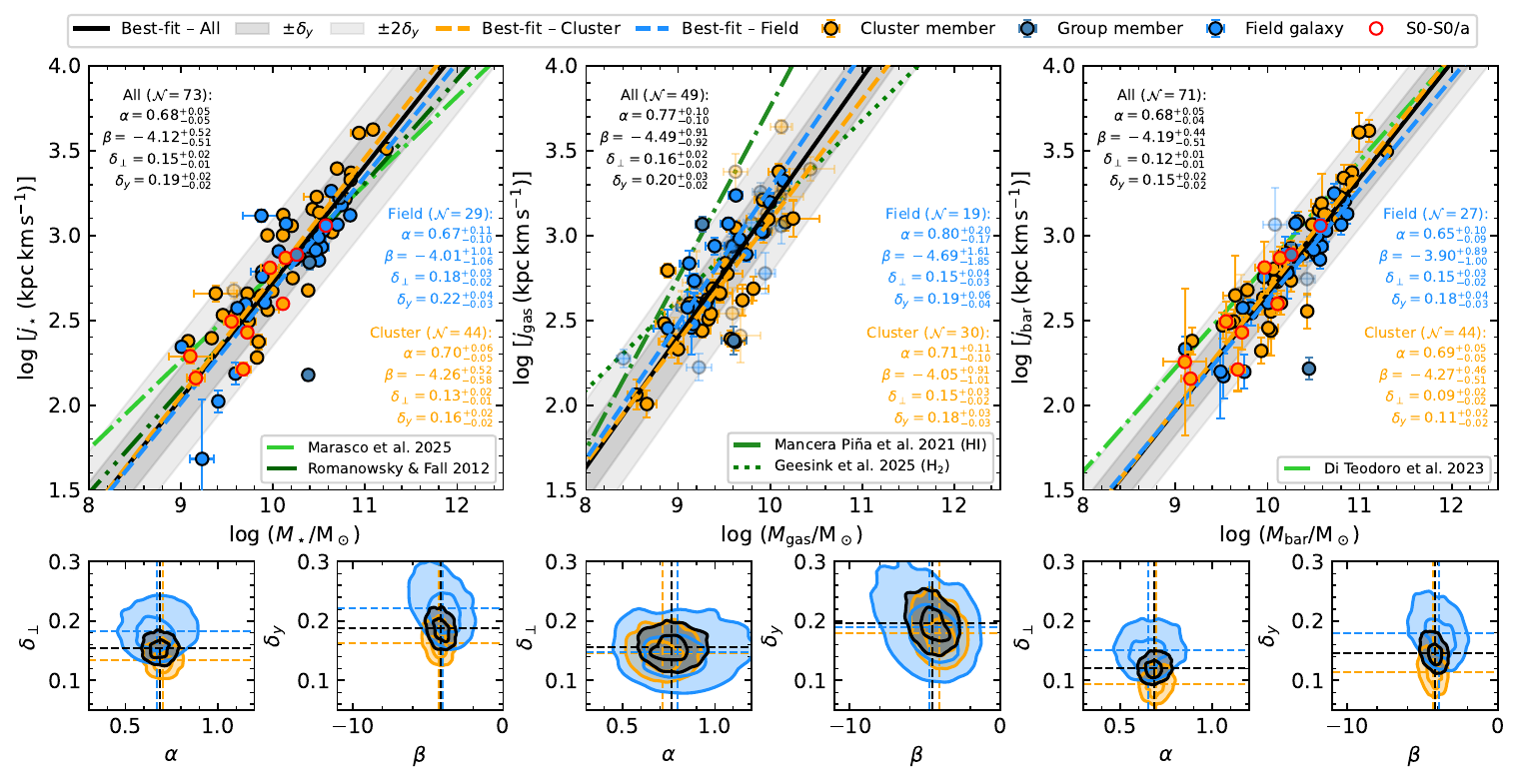}
	\caption{The SAMR for stars (left), gas (centre), and baryons (right) for cluster members and field galaxies, with the 2D posteriors of the best-fit parameters below each panel. 
	Shaded symbols mark galaxies with non converged $j$ (Appendix~\ref{app:j_convergence}). 
	Green lines are fits from the literature based on $V_\mathrm{rot,\star}$ \citep{2012RomanowskyFall}, $V_\mathrm{rot,\hi}$ \citep{2025Marasco,2023DiTeodoro}, $j_{\hi}$--$M_{\hi}$ \citep{2021ManceraPina}, and $j_\mathrm{H_2}$--$M_\mathrm{H_2}$ \citep{2025Geesink}.
	Only the gas SAMR shows a mild environmental dependence, with cluster members having lower $j_\mathrm{gas}$ than field galaxies. 
	}
	\label{fig:AMR}
\end{figure*}

As done for the TFRs, we test the environmental dependence of the SAMRs by fitting a power law to the fiducial sample and to the cluster and field sub-samples separately; the best fits are shown in Fig.~\ref{fig:AMR} and Table~\ref{tab:scaling_bestfit}. 
The stellar SAMR of the fiducial sample has $\alpha \approx 0.68$, $\beta \approx -4.12$, $\delta_\perp \approx 0.15$ and $\delta_y \approx 0.19$. 
The cluster and field best-fits are essentially identical, with largely overlapping 2D posteriors, indicating no obvious environmental dependence.
Over the common $M_\star$ range, cluster and field galaxies are offset by $\Delta_y = +0.10\pm0.05$~dex ($p_\mathrm{null} = 0.06$), consistent with zero and within $\delta_y$. 
The same holds for the baryonic SAMR, with $\alpha \approx 0.68$, $\beta \approx -4.19$, $\delta_\perp \approx 0.12$, and $\delta_y \approx 0.15$ for the fiducial sample, and $\Delta_y = +0.05\pm0.04$~dex between cluster and field galaxies. 
The stellar and baryonic parameters are very similar because our galaxies are relatively massive, so $j_\star$ and $M_\star$ dominate $j_\mathrm{bar}$ and $M_\mathrm{bar}$. 
Our results thus suggest that $j_\star$, and hence $j_\mathrm{bar}$, has not changed significantly since infall, and that most of its evolution precedes it. 
By the same argument pre-processing has a minor effect, at least for star forming and passive discs: we would see an imprint on $j_\star$ had it or $j_\mathrm{gas}$ been significantly modified before infall, for instance by ram pressure in filaments.
Our $\alpha \approx 0.65-0.70$ matches the DM SAMR predicted by tidal torque theory \citep{1980Fall,1983Fall}, suggesting these disc galaxies did not experience disruptive events (e.g. major mergers, strong AGN feedback). 
S0 and S0/a galaxies follow the same relation as other morphological types, both in clusters and in the field (see Sect.~\ref{sec:S0s} for discussion). 
Figure~\ref{fig:AMR_BT} shows that this holds also when the bulge is included.

The central panels of Fig.~\ref{fig:AMR} show that the gas SAMR of field galaxies is slightly steeper ($\alpha \approx 0.80$) than that of cluster members ($\alpha \approx 0.71$), with lower normalisation, although the differences are rather mild, below 1--2~$\delta$.
The gas SAMR is nonetheless the only relation in which the two sub-samples differ appreciably: over the common $M_\mathrm{gas}$ range, cluster members lie below field galaxies by $\Delta_y = -0.16\pm0.06$~dex ($p_\mathrm{null} = 0.008$).
This offset is not produced by modelling only one side of the disc in nine galaxies, since the model reproduces the unfitted side too (Sect.~\ref{sec:GasKin} and \citetalias{myGASPOnlineAtlas}). 
The \halpha\ discs of JO160, JO162, and JO194 seem genuinely shorter on the unfitted side, so $j_\mathrm{gas}$, integrated out to the more extended side, might be overestimated by $\lesssim 0.1-0.4$~dex. 
Since all three galaxies are stripped, this correction would increase the $j_\mathrm{gas}$ deficit. 
Its origin is further discussed in Sects.~\ref{sec:AMR_rps} and~\ref{sec:literature_AMR}, and quantified with a toy model in Appendix~\ref{app:toy}. 
The difference between the cluster and field $\alpha$ is exacerbated by the super-linear SK law used to convert \halpha\ into $\Sigma_\mathrm{gas}$, and shrinks further for the equivalent $j_\mathrm{H\alpha}$--$L_{\halpha}$ correlation, which avoids that conversion (Appendix~\ref{app:jhalpha}). 
Appendix~\ref{sec:AMR_phasespace} explores whether $j$ depends on the location in the phase-space diagram, indicative of infall time: galaxies with high $j$ are mostly extreme stripping (jellyfish galaxies), in the areas where recent infallers are most likely \citep[see also][]{2018Jaffe}. 

\subsubsection{Effect of ram pressure and AGN on the SAMRs}\label{sec:AMR_rps}
\begin{figure*}
	\centering
	\includegraphics[width=1\textwidth]{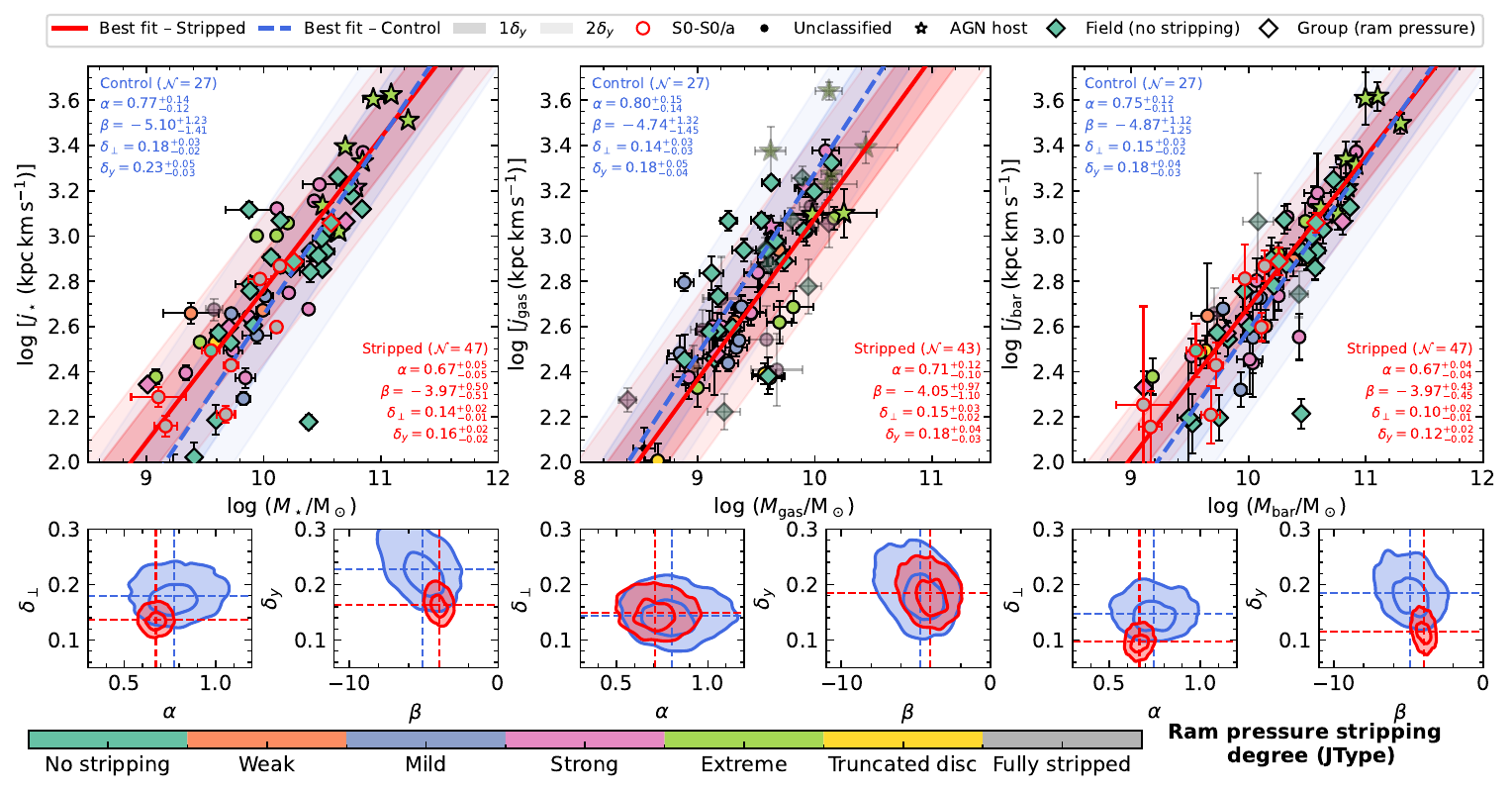}
	\caption{Same as Fig~\ref{fig:AMR} but with galaxies coloured according to their stripping stage.}
	\label{fig:AMR_JType}
\end{figure*}

Figure~\ref{fig:AMR_JType} shows the SAMRs coloured by degree of ram pressure stripping (JType), with AGN hosts marked by a different symbol. 
Unsurprisingly, the location on the stellar SAMR does not depend strongly on JType, since ram pressure affects the gas but not the stars. 
Jellyfish galaxies lie on average $\approx 0.2$~dex above the stellar SAMR, comparable to $\delta_y$, likely because they have marginally larger $M_\star$ and stellar discs than control galaxies \citep[Appendix A.1 in][]{2020Franchetto}. 
Our sample however is too small to establish whether this is a real feature or a selection effect. 
The relation of stripped galaxies is slightly flatter than that of the control sample, only because few control galaxies have low $M_\star$. 
In this plane, stripped and control galaxies differ, over the common $M_\star$ range, by $\Delta_y = +0.08\pm0.05$~dex ($p_\mathrm{null} = 0.14$).
Interestingly, extreme stripping galaxies lie above the relation, offset from the other galaxies by $\Delta_y = +0.17\pm0.07$~dex ($p_\mathrm{null} = 0.01$), that is $\approx \delta_y$. 
In the $j_\star$-$M_\star$ plane, AGN hosts show no offset ($\Delta_y = +0.05\pm0.07$~dex, $p_\mathrm{null} = 0.52$), although AGN feedback could in principle change $j_\star$ in either direction: ejecting low-$j$ gas raises $j_\star$ \citep{2012RomanowskyFall}, while suppressing the late-time accretion of high-$j$ gas lowers $j_\star$ ($\approx 0.3$~dex in case of early quenching, see \citealt{2017Lagos}; also \citealt{2015Genel,2022RodriguezGomez}). 
Being cumulative over a galaxy's history, these effects are not expected if the AGN was only recently triggered by ram pressure funnelling gas towards the centre \citep{2017PoggiantiNat,2023Akerman}. 

The most surprising feature of Fig.~\ref{fig:AMR_JType} is that a galaxy location in the $j_\mathrm{gas}$--$M_\mathrm{gas}$ plane does not depend on JType either, which seems counter-intuitive since $j_\mathrm{gas}$ should drop when gas is removed preferentially from the outskirts.
Over the common $M_\mathrm{gas}$ range, stripped galaxies lie below the control sample by $\Delta_y = -0.16\pm0.06$~dex ($p_\mathrm{null} = 0.008$).
The deficit shows no significant dependence on the degree of stripping, either in the gas plane ($\rho_\mathrm{Spear} = -0.26$, $p_\mathrm{Spear} = 0.17$) or when measured directly from \halpha\ ($\rho_\mathrm{Spear} = -0.23$, $p_\mathrm{Spear} = 0.20$; Appendix~\ref{app:jhalpha}).
This is because the offset produced by ram pressure saturates once the disc is truncated, since further stripping moves a galaxy along the relation rather than away from it.
The deficit is also found in the $j_\mathrm{H\alpha}$--$L_\mathrm{H\alpha}$ plane (Appendix~\ref{app:jhalpha}) and is thus not an artefact of the \halpha-to-gas conversion obtained by inverting the SK law.
The offset itself arises from the combination of the \halpha\ disc truncation and the $\approx 0.2$~dex enhancement in SFR \citep{2018Vulcani_b}, as discussed in Sect.~\ref{sec:literature_AMR} and Appendix~\ref{app:toy}.
\footnote{We cannot draw firm conclusions on the impact of AGN on the gas SAMR, as $\Sigma_\mathrm{gas}$ is assumed constant in those areas (see Sect.~\ref{sec:GasDist}). }

\section{Discussion}\label{sec:discussion}
We now compare our findings with literature on the TFRs and SAMRs (Sects.~\ref{sec:literature_TFR} and~\ref{sec:literature_AMR}), and discuss implications for S0's formation (Sect.~\ref{sec:S0s}) and evolution of scaling laws (Sect.~\ref{sec:evo_scaling_relations}). 

\subsection{Comparison with previous works on TFRs}\label{sec:literature_TFR}
This comparison is not always straightforward; previous studies used various velocity definitions, different methods for deriving $M_\star$ and the kinematics, and also kinematic tracers. 
Besides the $V_\mathrm{flat}$ used here, common velocity definitions are: $S_{0.5} = \sqrt{0.5V_\mathrm{rot}^2 + \sigma^2}$ \citep{2014Cortese}, the 21-cm integrated line profile width \citep[e.g.][]{1977TullyFisher,2023Catinella}, $V_\mathrm{circ}$ or $V_\mathrm{rot}$ measured at its peak ($V_\mathrm{max}$) or at a fixed radius ($R_\mathrm{e}$ or $2.2R_\mathrm{d}$). 
The last ones are the most used in IFU surveys with FOVs of $1-3 R_\mathrm{e}$, but are possibly affected by systematics. 
\cite{2019Lelli} recommend $V_\mathrm{flat}$ (Eq.~\ref{eq:Vflat}) since it minimises $\delta_\perp$ of the baryonic TFR based on $V_\mathrm{rot,\hi}$ from the Spitzer Photometry \& Accurate Rotation Curves (SPARC) sample, while other definitions are suboptimal. 
Our $V_\mathrm{circ}(R)$ are reasonably extended (median $R_\mathrm{max,V_{rot}} \approx 3.5 R_\mathrm{d}$; Fig.~\ref{fig:V_over_sigma}), reaching $2R_\mathrm{d}$, $3R_\mathrm{d}$, and $4R_\mathrm{d}$ for 95\%, 72\%, and 27\% of the sample, respectively, but still less extended than typical $V_\mathrm{rot,\hi}(R)$. 
Moreover, most \hi\ based studies infer $M_\star$ from the dynamical modelling of $V_\mathrm{circ,\hi}(R)$, using mid infrared (IR) luminosities and assuming motivated priors on the $\Upsilon_\mathrm{IR}$ \citep[e.g.][]{2016Lelli,2019Lelli,2021DiTeodoro_a,2023DiTeodoro}, while our $M_\star$ is from SED modelling on MUSE data. 

We therefore compare with \cite{2025Marasco} (see Fig.~\ref{fig:TFR}), who use $V_\mathrm{flat}$ from SPARC and $M_\star$ from SED modelling of ultra-violet to near-IR photometry, which is found to agree within $0.05$~dex with $M_\star$ from IR luminosity and dynamical modelling. 
They find $\alpha = 5.21 \pm 0.18$ for the stellar TFR, much steeper than ours. 
As shown in Appendix~\ref{app:M25}, this discrepancy is explained by two effects: the different $M_\star$ range covered by the samples, since we have few galaxies with $\log (M_\star/\mathrm{M_\odot}) \lesssim 9.5$, and the limited extent of our rotation curves \citep[see also][]{2023Catinella}.
Correcting for both, \citet{2025Marasco}'s sample gives $\alpha \approx 3.58$, $\beta \approx 2.48$ and $\delta_y \approx 0.24$ for a TFR based on the velocity at $2.2R_\mathrm{d}$, in agreement with our best fit. 

Our TFRs are broadly consistent with the $V_\mathrm{flat,\hi}$ derivations of \cite{2023DiTeodoro}, who add to the SPARC sample lower mass galaxies \citep{2021ManceraPina,2021ManceraPina_b} and super-massive spirals \citep{2021DiTeodoro_a}. 
These works assume $V_\mathrm{rot,\hi} \equiv V_\mathrm{circ}$ for the massive spirals, where $V_\mathrm{AD}$ is negligible, and correct for it in the dwarfs, as we do here; they also adopt $V_\mathrm{flat}$ as done here, but derive $M_\star$ from IR luminosity and dynamical modelling. 
Their baryonic $\alpha = 3.58\pm0.10$ agrees with ours ($3.48_{-0.20}^{+0.23}$) within $1\delta$, whereas their stellar $\alpha = 4.06\pm0.14$ is $\approx2\delta$ higher than our $3.58_{-0.21}^{+0.22}$. 
Considering our limited $M_\star$ range and rotation curve extents partially reconciles the different $\alpha$, but bring their slopes slightly below ours. 
This may hint at a bending at the highest masses, although \citet{2021DiTeodoro_a,2023DiTeodoro} find no statistical evidence for that. 
These \hi\ based studies build on samples representative of low density environments (although SPARC includes members of the diffuse cluster Ursa Major; \citealt{2001Verheijen}), supporting our finding that the TFRs (as defined here) are environment independent.

Our results agree with studies based on H$\alpha$ kinematics reporting no strong environmental dependence, but the comparison is less direct because they use $V_\mathrm{max}$ \citep{2013TorresFlores} or $V_\mathrm{rot}$ at $1-2R_\mathrm{e}$ \citep{2024Ristea}. 
\cite{2013TorresFlores} find no difference between the TFRs of field galaxies in the Gassendi HAlpha survey of SPirals \citep[GHASP;][]{2011TorresFlores} and in the Hickson Compact Groups, where tidal interactions should be significant. 
Using MaNGA, \citet{2024Ristea} derive the stellar TFR over a mass range similar to ours, finding that its scatter correlates strongly with $V_\mathrm{rot}/\sigma$ but not with environment. 
We do not find such correlations, likely because we correct for the asymmetric drift, while \citet{2024Ristea} do not. 

\subsection{Comparison with previous works on SAMRs}\label{sec:literature_AMR}
Most IFU-based studies use the spin parameter $\lambda_\star = \langle R |V_\mathrm{rot,\star}|\rangle / \langle R \sqrt{V_\mathrm{rot,\star}^2 +\sigma_\star^2}\rangle$ \citep{2007Emsellem}, the brackets denoting a flux weighted average over the pixels. 
Like $V_\mathrm{rot}/\sigma$, $\lambda_\star$ roughly quantifies the rotation dominance in a region and is thus different from $j$. 
In the following, unless specified, we compare with studies using the same $j$ definition as here (Eq.~\ref{eq:j_1comp}), which refers to the disc alone. 
Figure~\ref{fig:AMR} compares our stellar SAMR with those for discs from \citet{2012RomanowskyFall}, who measure $j_\star$ from $V_\mathrm{rot,\star}$ and $M_\star$ from $K$ band photometry, and \citet{2025Marasco}, based on $V_\mathrm{rot,\hi}(R)$ from SPARC and $M_\star$ from SED fitting. 
At the median $M_\star$ of our sample, the relations agree within their same scatter ($\delta_y \approx 0.17-0.19$). 
\citet{2012RomanowskyFall} measure $\alpha = 0.61\pm0.04$, which is within $\approx1\delta$ from ours, whereas \citet{2025Marasco} as well as other determinations for discs based on $V_\mathrm{rot,\hi}(R)$ are $\approx3\delta$ flatter ($\alpha \approx 0.52-0.55$; \citealt{2018Posti,2021ManceraPina,2023DiTeodoro,2025Marasco}). 
The range of $M_\star$ and how it is measured cannot explain this difference, which could perhaps result from the different radial extent over which $j_\star$ is measured, or the inference of $V_\mathrm{rot,\star}$ from $V_\mathrm{rot,\hi}$. 
Besides these small differences, our stellar SAMR agrees with other determinations in the literature for discs in low-density environments, supporting our finding of a lack of environmental dependence.
This agrees also with \citet{2021Mostoghiu}, who follow the evolution of $j_\star$ (although for the bulge and disc together) for satellites with $M_\star > 5 \times 10^{10}$~\msun\ in the \textsc{Three Hundred Project} simulations. 
They find $j_\star$ is preserved after infall, the stellar kinematics being unchanged even though DM and gas are stripped. 
\citet{2024Croom} reach a similar conclusion from the SAMI survey and the \textsc{Eagle} simulations, although based on $\lambda_\star(<R_\mathrm{e})$ \citep[see also][]{2019Cortese}. 

What was said above also applies to the baryonic SAMR: its slope and scatter agree with the literature \citep[e.g.][Fig.~\ref{fig:AMR}]{2023DiTeodoro} and with the stellar SAMR, as expected since stars dominate $j_\mathrm{bar}$ and $M_\mathrm{bar}$ in most of our sample.
The lack of strong environmental dependence supports the conclusions in \cite{2020Murugeshan}, who find no strong trend between local density and $j_\mathrm{bar}$. 
Conversely, \cite{2024Sorgho} report that isolated discs have larger $j_\mathrm{bar}$ than non-isolated ones, though the difference is arguably minor compared to $\delta_y$. 

The gas SAMR is harder to interpret than the stellar and baryonic ones. 
We find $\alpha \approx 0.77$ for the whole sample, $\alpha \approx 0.80$ for field galaxies, and $\alpha \approx 0.71$ for cluster members. 
This difference is exacerbated by the super-linear SK law, being milder for the relation between $j_\mathrm{H\alpha}$ and $L_\mathrm{H\alpha}$ (Appendix~\ref{app:jhalpha}).
Nevertheless, stripped galaxies lie $\approx 0.16$~dex below the relation for unperturbed ones, a difference that persists for the $j_\mathrm{H\alpha}$--$L_\mathrm{H\alpha}$ correlation. 
The offset has two possible origins. 
The first is a lower $V_\mathrm{rot, \hii}$ at fixed mass in stripped galaxies, but Fig.~\ref{fig:TFR_JType} shows that their $V_\mathrm{flat,\hii}$ is similar to the control one, and the marginal negative trend of $(V_\mathrm{rot}/\sigma)_\mathrm{\hii}$ with JType (Appendix~\ref{app:Vsigma_rps}) is not mirrored by $V_\mathrm{rot, \hii}$. 
The second is that stripped galaxies have more centrally concentrated \halpha\ discs, being smaller but more star forming \citep{2011Jaffe,2018Vulcani_b}; Appendix~\ref{app:toy} shows that this alone can explain the offset. 
Indeed $V_\mathrm{rot,\hii}$ is measured on average at smaller $R$ in cluster than in field galaxies (Fig.~\ref{fig:V_over_sigma}), and likewise in stripped than in control ones (Appendix~\ref{app:toy}). 
Figure~\ref{fig:AMR} shows that our best fit lies $\approx 0.2$--$0.8$~dex below the linear $j_\mathrm{\hi}$--$M_\mathrm{\hi}$ relation of \cite{2021ManceraPina}. 
Our best fit is $\approx2\delta$ steeper than the H$_2$ relation of \citet{2025Geesink} ($\alpha \approx 0.77$ vs $0.53$) but only marginally offset ($\lesssim0.3$~dex) over our $M_\mathrm{gas}$ range.
The comparisons are not direct, however, since our $\Sigma_\mathrm{gas}$ is inferred from \halpha\ through the SK law, rather than measured from spatially-resolved \hi\ and CO observations. 
We conclude that \halpha\, which traces the inner star-forming disc, misses part of the total $j_\mathrm{gas}$, being less accurate than \hi\ but comparable to CO.

Our findings indirectly constrain the origin of the \halpha\ tails. 
Simulations show that ram pressure perturbs the gas rotation curve, mostly in the outer disc \citep{2008Kronberger,2023Akerman}, so tails made of gas displaced from the disc would change both $V_\mathrm{flat}$ and $j$. 
For the baryonic TFR, we measure $\Delta_y = -0.06\pm0.07$~dex and $\delta_y \approx 0.22$~dex, which imply that any reduction of the outer $V_\mathrm{circ}$ is $\lesssim 14$\%. 
Such a reduction at $R>R_\mathrm{d}$ would produce a difference in $\log j$ of $\lesssim 0.05$~dex, insufficient to explain the observed offset without the SFR enhancement (see Appendix~\ref{app:toy}). 
The tails must then have a different origin: most likely, \hi, being distributed in more extended discs than \halpha, was stripped and then ionised by star formation in situ. 
Previous studies indeed find star forming clumps with \hii-region line ratios in the tails of jellyfish galaxies \citep[e.g.][]{2019Poggianti} and truncated \hi\ discs \citep[e.g.][]{2019Ramatsoku,2020Ramatsoku}.

\subsection{Formation scenarios of S0s}\label{sec:S0s}
There are two main channels for the formation of S0s, each predicting different $j_\star$. 
In the faded spiral scenario, S0s descend from spirals that consumed or lost their gas reservoir \citep[e.g.][]{2017Bellstedt,2018Rizzo,2023Marasco_b,2023Vulcani_c}. 
In the merger scenario, S0s form via mergers that trigger starbursts, causing a rapid gas consumption and the bulge growth \citep[e.g.][]{1998Bekki,2015Querejeta}. 
The faded spiral scenario predicts that $j_\star$ is conserved, while mergers are expected to reduce it \citep[e.g.][]{2017Lagos}. 
Figure~\ref{fig:AMR} shows that the discs of our S0--S0/as closely follow the same stellar SAMR as star-forming discs ($\Delta_y = -0.03\pm0.07$~dex, $p_\mathrm{null} = 0.67$), suggesting that they are most likely descendants of spirals quenched by ram pressure or starvation, without disruptive events that would redistribute $j_\star$. 
\cite{2018Rizzo} analyse ten field S0s from the Calar Alto Legacy Integral Field Area (CALIFA) survey. 
They use the same $j_\star$ definition used here but separate bulge and disc both photometrically (as we do) and kinematically (not done here). 
Considered together, our S0--S0/as and those from \citet{2018Rizzo} cover the range $9 \lesssim \log (M_\star/\mathrm{M_\odot}) \lesssim 11.5$, reaching the high $M_\star$ regime where mergers are thought to dominate \citep[e.g.][]{2018FraserMcKelvie,2021Pulsoni,2026Pallero}. 
Eight S0s from \cite{2018Rizzo} are fully consistent with our stellar SAMR, with the only two outliers showing signs of past mergers; this suggests that the faded spiral scenario may be the main formation channel regardless of environment \citep[see also][]{2022Coccato}. 
Studies based on $\lambda_\star$ \citep[e.g.][]{2015Querejeta} or the disc-plus-bulge $j_\star$ \citep[e.g.][]{2023Pulsoni} find S0s $\approx 0.6$~dex below the spirals' SAMR, likely because $\lambda_\star$ is a different quantity than $j_\star$ and including the bulge lowers $j_\star$. 
In fact, bulges tend to be more pressure-supported than discs and likely follow a different formation pathway. 
Nevertheless, including bulges, we find just a marginally shallower relation (Appendix~\ref{app:j_bulge}), likely because few galaxies have $B/T>0.3$; only JW105 moves $\approx 2\delta_y$ from the relation.

Figure~\ref{fig:TFR} shows that our S0--S0/as follow the TFRs as closely as star-forming discs. 
This corroborates studies showing that the offset of S0s from the luminosity--velocity TFR of spirals \citep[e.g.][]{2013Rawle,2014Jaffe} reflects their different star formation history, and hence $\Upsilon$, rather than a change in mass or velocity \citep[e.g.][]{2015DenHeijer}.
In fact, other GASP studies find a rapid, environment-driven transformation: \citet{2020Vulcani_a} find signs of abrupt quenching in our S0--S0/as and \citet{2023Marasco_b,2026Marasco} show that stellar population ageing alone turns quenched spirals into S0s in $\approx 1.5-3.5$~Gyr, reproducing the passive disc population in clusters.

\subsection{Evolution of scaling relation with $z$}\label{sec:evo_scaling_relations}
Being based on optical observations, our relations could be used as a $z \sim 0$ reference\footnote{This assumes that our sample is representative of present-day discs, which seems reasonable since our stellar and baryonic scaling relations broadly agree with those for larger samples at low redshift.} for studies of their evolution with cosmic time, mostly based on ionised gas tracers (e.g. \halpha, [OII]; \citealt{2019Marasco,2021PerezMartinez,2023Mercier,2025EspejoSalcedo,2026ManceraPina}) and, rarely, stars \citep[e.g.][]{2024MunozLopez,2025Slob}. 
The closest to ours is \cite{2026ManceraPina}: they measure \halpha\ kinematics with \barolo\ over $5 \times 10^{9} \lesssim M_\star/\mathrm{M_\odot} \lesssim 2 \times 10^{11}$, adopt the same $V_\mathrm{flat}$ definition and asymmetric drift correction, and derive $M_\star$ from SED modelling. 
Our stellar TFR is fully consistent with theirs at $z \approx 0.9$ ($\approx 8$~Gyr ago), which has $\alpha = 3.82^{+0.55}_{-0.40}$ and $\delta_\perp = 0.06 \pm 0.01$ ($\delta_y \approx 0.26$), similarly to previous studies using \halpha\ IFU data \citep[e.g.][]{2016DiTeodoro,2017Ubler,2019Tiley}. 
Section~\ref{sec:literature_TFR} shows that the differences between our stellar TFR and the \hi-based ones at $z \approx 0$ come from the different $M_\star$ range and the limited extent of $V_\mathrm{rot,\hii}(R)$, so part of the moderate evolution reported by \cite{2026ManceraPina} may reflect these differences rather than a genuine change with $z$. 
On the other hand, our stellar SAMR is steeper than \cite{2026ManceraPina}'s one, which has $\alpha = 0.44 \pm 0.06$: the two cross at $\log (M_\star/\mathrm{M_\odot}) \approx 9.8$ but ours is $\approx 0.4$~dex above at high $M_\star$. 
This supports their conclusion that the stellar SAMR has evolved in the past $\approx 8$~Gyr, mostly in zero-point rather than $\alpha$, at least when ionised gas is used as kinematic tracer. 
We recall that using ionised gas tracers to probe the evolution of scaling relations is risky, yet unavoidable given the integration time to detect cold gas tracers at high $z$. 
Non-circular motions \citep[e.g.][]{2021Girard,2022Ejdetjarn,2024Kohandel,2024Rizzo} and beam smearing are harder to correct at coarser resolution, introducing systematics in the scaling relations.
These issues are milder here: MUSE resolves our discs with more than 10 resolution elements along the major axis, against five at high $z$, and our $V_\mathrm{rot,\hii}(R)$ reach $20-30$~kpc, against $8-10$~kpc \citep[e.g.][]{2016DiTeodoro,2026ManceraPina}.

\section{Conclusions}\label{sec:conclusions}
This study investigates the impact of the large scale environment on the Tully-Fisher relations (TFRs) and specific angular momentum - mass relations (SAMRs) of disc galaxies, using MUSE data from the GASP survey \citep{2017Poggianti}, which covers cluster members and galaxies in low density environments (groups and the field) at $z \approx 0.05$. 
The parent sample has more than 100 galaxies; $\approx 10$\% are passive discs (S0--S0/a), the rest are star-forming spirals.
We constrain the disc geometry from $I$~band images, the stellar distributions and masses from SED modelling, and the cold gas distribution and mass from the \halpha\ emission by inverting the Schmidt-Kennicutt law, using \hi\ and CO observations when available.
We measure the stellar and ionised gas kinematics at $\approx 1$~kpc resolution, apply the asymmetric drift correction to obtain $V_\mathrm{circ}$ and hence $V_\mathrm{flat}$, and compute the specific angular momenta of stars ($j_\star$), gas ($j_\mathrm{gas}$), and baryons ($j_\mathrm{bar}$), searching for systematic differences as a function of environment and degree of ram pressure stripping.
Our main results are the following. 
\begin{itemize}
	\item 
	The stellar and ionised gas kinematics ($V_\mathrm{rot}$, $\sigma$, and $V_\mathrm{rot}/\sigma$) show no significant dependence on environment. 
	$(V_\mathrm{rot}/\sigma)_\mathrm{\hii}$ decreases mildly with the degree of ram pressure stripping, but the trend is not statistically significant. 
	
	\item 
	Cluster and field galaxies follow the same TFRs, regardless of the kinematic tracer (stars or ionised gas), even under extreme ram pressure stripping (jellyfish galaxies). 
	This suggests that any change in their baryonic content, kinematics, and gravitational potential is too small to produce detectable deviations with the disc area analysed here. 
	
	\item 
	The stellar and baryonic SAMRs do not strongly depend on environment or degree of stripping, so neither cluster infall nor pre-processing has modified $j_\star$ and $j_\mathrm{bar}$ beyond the scatter of the relations. 
	
	\item 
	The gas SAMR of field galaxies is slightly steeper than that of cluster members ($\alpha \approx 0.80$ against $0.71$), which have smaller, more centrally concentrated \halpha\ discs and lie $\approx 0.16$~dex below. 
	The deficit does not depend on the degree of stripping, jellyfish galaxies included, and is mostly due to the SFR enhancement of galaxies affected by ram pressure. 
	
	\item 
	S0 and S0/a galaxies follow the same stellar SAMR as late-type discs, in clusters and in the field alike, consistent with them being faded spirals. 
	
\end{itemize} 
Our results fit the picture in which the ionised tails of galaxies affected by ram pressure originate from neutral gas stripped from the ISM and ionised by in-situ star formation. 
At low redshift, the cluster environment quenches star-forming galaxies mainly by removing the ISM and CGM (thus preventing accretion) through ram pressure, while the disc $j_\star$ stays largely intact because most of its evolution precedes infall and pre-processing in filaments did not produce detectable changes. 


\begin{acknowledgements} 
The authors thank Eric Emsellem and Claudia Pulsoni for helpful comments and feedback on this work, and Yara Jaff\'{e} and Amir H. Khoram for useful comments on the paper. 
CB acknowledges support from the Carlsberg Foundation Fellowship Programme by Carlsbergfondet. 
PEMP is funded by the Dutch Research Council (NWO) through the Veni grant VI.Veni.222.364. 
This research made use of artificial intelligence models (Claude by Anthropic) for improving codes and text. 
\end{acknowledgements}

\bibliographystyle{aa}
\bibliography{paty.bib}

\begin{appendix}

\section{Galaxy sample and atlas description}\label{app:atlas}
Table~\ref{tab:sample} reports the GASP ID for the parent sample, the environment, and the morphological classification, followed by the logarithmic quantities for the scaling relations studied here ($M$, $V_\mathrm{flat}$, and $j$).
Based on MUSE observations (see Sect.~\ref{sec:class}), we update the morphological class of some galaxies. 
We reclassify JO24 and P4946 as Sa rather than S0, since the MUSE revealed an \halpha\ disc, JW105 as S0/a because it has faint spiral arms, and A3158\_B\_0223 as a barred S0 rather than Sa. 
Our S0/a sample also includes "anaemic spirals", passive discs with spiral arms and bulges (A3158\_11\_91, A3376\_B\_0214, A3158\_B\_0234, and P5169), and two passive galaxies of unclear morphology (A500\_F\_0152 and JW36), classified as $k+a$ post starburst affected by ram pressure stripping \citep{2020Vulcani_a}. 
Star-forming discs (Sa to Scd) are treated together, so no revision is needed.

\begin{table*}
\setlength{\tabcolsep}{2pt}
\footnotesize
\caption{Properties of the parent sample from GASP: Environment (Env.) with C\,=\,cluster, G\,=\,group and F\,=\,field, morphological type (Morph.), ram pressure stripping degree (JType), stellar disc mass ($\log M_\mathrm{\star,d}$), total stellar mass ($\log M_\mathrm{\star}$), cold gas mass ($\log M_\mathrm{gas}$), baryonic mass ($\log M_\mathrm{bar}$), average velocity in the flat or outermost part of the circular velocity curve ($\log V_\mathrm{flat}$), specific angular momentum of the disc of stars ($\log j_\mathrm{\star}$), gas ($\log j_\mathrm{gas}$), and baryons (gas+stars) ($\log j_\mathrm{bar}$). Daggers ($\dagger$) mark galaxies whose $M_\mathrm{gas}$ is rescaled to match the observed $M_\hi$ or $M_\hi$+$M_\mathrm{H_2}$.}
\label{tab:sample}
\centering
\begin{tabular}{lccccccccccc}
\hline\hline
Galaxy & Env. & Morph. & JType & $\log M_{\star,d}$ & $\log M_\star$ & $\log M_\mathrm{gas}$ & $\log M_\mathrm{bar}$ & $\log V_\mathrm{flat}$ & $\log j_{\star}$ & $\log j_\mathrm{gas}$ & $\log j_\mathrm{bar}$ \\
 & & & & [$M_\odot$] & [$M_\odot$] & [$M_\odot$] & [$M_\odot$] & [km\,s$^{-1}$] & [kpc\,km\,s$^{-1}$] & [kpc\,km\,s$^{-1}$] & [kpc\,km\,s$^{-1}$] \\
\hline
A1069\_B\_0103 & C & S0 & 4.0 & $9.97 \pm 0.10$ & $10.21 \pm 0.15$ & \ldots & $10.21 \pm 0.15$ & $2.25 \pm 0.03$ & $2.81 \pm 0.02$ & \ldots & $2.81 \pm 0.15$ \\
A3128\_B\_0148 & C & Sab & 0.5 & $9.83 \pm 0.06$ & $9.83 \pm 0.06$ & $9.41 \pm 0.13$ & $9.97 \pm 0.06$ & $2.10 \pm 0.04$ & $2.28 \pm 0.03$ & $2.44 \pm 0.04$ & $2.33 \pm 0.08$ \\	
\hline
\end{tabular}
\tablefoot{Quantities are omitted when a galaxy is excluded from the analysis (the motivation is reported), is passive (no \halpha, so no $M_\mathrm{gas}$ and $j_\mathrm{gas}$), or the kinematics of stars and/or ionised gas is unreliable (no $j$); ">" is written before non-converged $j$. 
The entire table is available \href{https://doi.org/10.5281/zenodo.23061246}{online}.}
\end{table*}
\normalsize

We provide an online \citetalias{myGASPOnlineAtlas} with our sample, data and models. 
Each page shows one galaxy, with the title reporting GASP ID, redshift $z$, stripping class (JType), and morphology; the 24 panels have titles in gold for photometry and stellar distribution, orange for stellar kinematics, and red or purple for the ionised gas. 
Cases with unreliable kinematics are shown in greyscale. 

\begin{enumerate}[A)]
	
	\item $I$ band image with the mask overlaid (white). 
	Grey contours trace isophotes at $n \times I_\mathrm{sky}$, with $n = 1, 2, \ldots 10$. 
	The magenta dashed ellipse is the outermost fitted isophote with S/N>1, radius $R_\mathrm{max}$, and photometric centre $(x_0, y_0)$ (black cross). 
	
	\item Surface brightness profile $I(R)$ (black points). 
	The solid red curve is the best-fit model, consisting of an exponential disc (orange dashed line) plus S\'{e}rsic bulge (golden dotted curve), when needed.
	The downward triangle points to $R_\mathrm{max}$. 
	The grey dashed curve show is the cumulative $I(R)$ in arbitrary units.
	The teal dotted profile is obtained from $\Sigma_\star$ model and $\Upsilon^I$. 
	The bottom panel shows the residuals $\Delta\mathrm{SB} = \mathrm{data} - \mathrm{model}$ in mag~arcsec$^{-2}$; the red dashed line is $\Delta\mathrm{SB}=0$.
	
	\item Stellar surface mass density ($\Sigma_\star$) map from SED fitting. 
	The black cross shows $(x_0, y_0)$. 
	
	\item Stellar surface density radial profile $\Sigma_\star(R)$ (black points). 
	The solid red, dashed orange, and dotted gold curves are the best-fit models for the total $\Sigma_\star(R)$, the disc, and bulge, respectively. 
	For each component, the legend reports the model parameters, mass-to-light ratios $\Upsilon^I$, masses, and $B/T$. 
	The downward triangles in orange and grey mark, respectively, the disc scale length $R_\mathrm{d}$ and the bulge effective radius $R_\mathrm{e}$. 
	The bottom panel displays the residuals $\Delta\log\Sigma_\star$ in dex; the red line is $\Delta\log\Sigma_\star=0$.
	
	\item Radial profile of $x_0$ from isophote fitting (black points) and adopted centre $x_0$ (orange dashed line).
	
	\item Same as Panel~E but for the centre coordinate $y_0$. 
	
	\item Radial profile of the ellipse inclination from the fitted isophote $\epsilon$, with the adopted $i$ (orange dashed line).
	
	\item Same as panel~G but for the ellipse position angle $\phi$. 
	
	\item Stellar velocity field ($V_\mathrm{LOS}-V_\mathrm{sys}$).
	The black contour is at zero. 
	The white dashed line and ellipse indicate the kinematic major axis and the outermost fitted ring by the kinematic model. 
	The black cross corresponds to the centre. 
	The black circle (bottom left corner) and the bar (top right corner) show, respectively the observational seeing of 1\arcsec and the physical scale corresponding to 5~kpc.
	
	\item Model of the stellar velocity field.
	
	\item Residual map: stellar velocity field minus the model. 
	
	\item Stellar velocity dispersion map. 
	
	\item Radial profiles of $i$ (top) and $\phi$ (bottom) of the best-fit model in panel J. 
	Grey points are individual ring values from the second fit; 
	Orange dashed curves are the regularised profiles used in the final fit.
	
	\item Radial profiles of rotation velocity $V_\mathrm{rot,\star}$ from the model in panel J (open stars), asymmetric drift velocity $V_\mathrm{AD,\star}$ (small filled stars), and circular velocity $V_\mathrm{circ,\star}$ (large filled stars). 
	
	\item Radial profile of the stellar velocity dispersion $\sigma_\star$ (stars) from the map in panel L.
	The black dashed curve shows the exponential model used for regularising $\sigma_\star(R)$ in the computation of $V_\mathrm{AD,\star}(R)$.
	
	\item Comparison of $V_\mathrm{circ}(R)$ for stars (orange stars) and ionised gas (red circles). 
	The black crosses show the fiducial $V_\mathrm{circ}(R)$. 
	
	\item Ionised gas total intensity map (0th moment of the \halpha\ or \hbeta\ datacube, whichever is the best kinematic tracer). 
	The white cross, black ellipse and dashed line show the kinematic centre, the outermost fitted ring and the kinematic major axis. 
	
	\item Position--velocity diagram along the kinematic major axis extracted from the ionised gas datacube (red for \halpha, purple for \hbeta). 
	Black contours are at $2^n\,\delta_\mathrm{ch}$ ($n = 0, 1, \ldots 10$), where $\delta_\mathrm{ch}$ is the noise per channel; grey contours are at $-\delta_\mathrm{ch}$. 
	The red contours show the best-fit model, at the same levels as the observations. 
	Yellow points track the model rotation curve. 
	The horizontal and vertical dashed lines correspond to $V_\mathrm{sys}$ and the kinematic centre.
	When present, the shaded grey vertical band covers the radii excluded from the fit because of AGN or \nii\ contamination; when a disc side is not used in the fit, its quadrants are cross-hatched.
	
	\item Radial profiles of $V_\mathrm{rot,\hii}$ (open circles), $V_\mathrm{AD,\hii}$ (small filled circles), and $V_\mathrm{circ,\hii}$ (large filled circles). 
	
	\item Radial profile of the cold gas surface density $\Sigma_\mathrm{gas}(R)$ (red circles), derived from the azimuthally-averaged \halpha\ surface brightness profile (Sect.~\ref{sec:GasDist}). 
	When present, the shaded grey vertical band covers the radii excluded from the fit (item R), and the points with a white edge are extrapolated assuming $\Sigma_\mathrm{gas}$ constant at the innermost measured value. 
	
	\item Ionised gas velocity field (1st moment of the datacube, minus $V_\mathrm{sys}$). 
	The thick black contour shows the isovelocity curve at zero; solid and dashed thin contours are positive and negative isovelocity curves spaced by 50~\kms. 
	The white cross, black circle (bottom left corner), and black bar (top right corner) show the kinematic centre, the seeing of 1\arcsec, and a physical scale of 5~kpc, respectively. 
	
	\item Same as panel R but along the minor kinematic axis. 
	
	\item Radial profile of the ionised gas velocity dispersion $\sigma_\mathrm{\hii}(R)$ (red circles). 
	The black dashed curve is the exponential model used for regularising the profile in $V_\mathrm{AD,\hii}$ computation.
	
	\item Radial profiles of $i$ (top) and the kinematic $\phi$ (bottom) from \barolo model of the ionised gas datacube. 
	Grey points are individual ring values from the first fit; red curves are the regularised profile in the final fit. 
	
\end{enumerate}

\section{More on fitting procedures}\label{app:fitting}
This section clarifies the fitting procedures for photometric images, emission line datacubes, stellar densities, and scaling relations, as well as the computation of uncertainties. 

\subsection{Isophote fitting and disc geometry}\label{app:isophotes}
Before the isophote fitting (Sect.~\ref{sec:Geometry}), $I_\mathrm{sky}$ is subtracted from the image following a procedure similar to that used in \cite{2023Marasco}. 
The sky-dominated image is modelled with a Gaussian plus Schechter function: the mean and standard deviation of the Gaussian give $I_\mathrm{sky}$ and its rms $\delta I_\mathrm{sky}$, while the Schechter term accounts for the high-$I$ tail from point-like sources and resolved systems.
Each ellipse is spaced by three pixels ($0.6$\arcsec, $\approx$ the Nyquist sampling).
The inclination $i$ and its uncertainty $\delta i$ are derived via $(\cos i)^2 = \left[(b/a)^2 - q_0^2\right]/(1-q_0^2)$, with intrinsic axis ratio $q_0=0.13$ \citep{2008Padilla}.
The galaxy centre is determined as the median $x_0$ and $y_0$ within $R \leq 5$~kpc, while the median $i$ and $\phi$ are computed for $R$ larger than 3.5~kpc or the bar length from \cite{2023SanchezGarcia}, whichever is the largest, so as to reduce the contamination from bars and bulges.
Uncertainties on the geometrical parameters are the semi-difference between the 84th and 16th percentiles of the ellipse parameters.

For each ellipse we compute the mean intensity $I(R)$ and its uncertainty $\delta I(R)$, the larger of the pixel standard deviation and $\delta I_\mathrm{sky}$, divided by the square root of the number of independent resolution elements in the ring.
We extract $I(R)$ in progressively larger rings out to $R_\mathrm{max}$, defined as the radius where $I(R)/\max(\delta_\mathrm{rms},\delta I_\mathrm{sky}) < 1$, where $\delta_\mathrm{rms}$ is the rms in the ring normalised by the square root of the number of sampled pixels. 
Foreground stars and other galaxies are masked with the \texttt{segmentation} tool, and within each ellipse pixels deviating by more than $3\delta I_\mathrm{sky}$ are clipped, further reducing the contamination from unmasked sources. 
The cumulative light profile flattens by the outermost ellipse, confirming that $I_\mathrm{sky}$ is well estimated and no flux is left beyond $R_\mathrm{max}$.

\subsection{Uncertainties and error propagation}\label{app:errors}
Unless stated otherwise, uncertainties are propagated with a Monte Carlo scheme: we draw $N=10^4$ realisations of the input quantities from Gaussians centred on their best-fit or measured values, with dispersion given by their uncertainties, and take the median and the semi-difference of the 84th and 16th percentiles.

In each ring of the stellar mass maps (Sect.~\ref{sec:StarDist}), $\delta \Sigma_\star$ is the standard deviation of the pixels with measured stellar mass divided by the square root of the number of independent resolution elements.
We stop at the outermost ring where at least 80\% of the pixels have a measured stellar mass, keeping below $0.1$~dex the error from pixels with S/N under the SED fitting threshold.

For the gas surface densities (Sect.~\ref{sec:GasDist}), each realisation draws $a$ and $A$ from Gaussians centred on the best-fit values of \cite{2021Kennicutt}, and rescales $\Sigma_\mathrm{H\alpha,corr}(R)$ so that its disc average matches the $\Sigma_\mathrm{gas}$ from Eq.~\ref{eq:kslaw}; $M_\mathrm{gas}$ and its uncertainty $\delta M_\mathrm{gas}$ follow from the same realisations. 

In Sect.~\ref{sec:StarKin}, for each annulus of the best-fit model of the stellar velocity field, $\delta V_\mathrm{rot,\star}$ sums in quadrature the inclination uncertainty $\delta i$ and the statistical uncertainty on the mean $V_\mathrm{rot,\star}$: 
\begin{equation}
 \delta V_\mathrm{rot,\star} = \sqrt{
			\left(\frac{V_\mathrm{rot,\star}\delta i}{\tan i}\right)^2 + 
			\left(\frac{\delta V_\mathrm{LOS,res}}{(\sin i \sqrt{\sum_b \cos^2\theta_b})} \right)^2 }\, \,
\end{equation}
where $\delta V_\mathrm{LOS,res}$ is $1.48\times$ the median absolute deviation (MAD) of the residual in the ring computed along independent resolution elements (i.e. the bins in the outer disc where the bins are larger than the PSF, or the PSF area when this is larger than the bins), the sum $\sum_b \cos^2\theta_b$ runs over the bins $b$ in the ring, $\theta_b$ is the bin angle measured from the major axis, and $\cos^2\theta_b$ is the weighting used in the tilted-ring fit.
The latter is rescaled by the number of resolution elements per bin, whenever a ring contains more bins than resolution elements.

For the circular velocities (Sect.~\ref{sec:VAD_Vcirc}), each realisation of Eqs.~\ref{eq:vc_vad} and~\ref{eq:vad} draws from the models for $\Sigma(R)$ and $\sigma(R)$ and from the observed $V_\mathrm{rot}$.
The uncertainty on $V_\mathrm{flat}$ is computed as 
\begin{equation}\label{eq:Vflat}
	\delta {V_\mathrm{flat}} = \sqrt{\frac{1}{N} \sum_{i}^{N} \delta{V_\mathrm{circ,i}^2} + \delta V_\mathrm{ave}^2} \,;
\end{equation}
where $\delta {V_\mathrm{circ,i}}$ and $\delta V_\mathrm{ave}$ are respectively the uncertainties on each circular velocity point and the standard deviation around $V_\mathrm{ave}$. 
For each $j$ component (Sect.~\ref{sec:AMR}), the realisations sample the stellar disc parameters, $\Sigma_\mathrm{gas}$, and $V_\mathrm{rot}$.

\subsection{Bulge and disc models}\label{app:bulge_disc}
We separate disc and bulge using both $I(R)$ and $\Sigma_\star(R)$. 
Since the S/N cut of the SED fitting makes $I(R)$ reach larger $R$ than $\Sigma_\star(R)$, we use $I(R)$ to probe the outskirts and constrain $R_\mathrm{d}$. 
With \textsc{dynesty} \citep{2020Speagle,2023Koposov} for nested sampling \citep{2004Skilling,2006Skilling}, we fit $I(R< R_\mathrm{max})$ with two models, adding an intrinsic scatter $\delta_\mathrm{int}$ for the uncertainties not included in $\delta I$.
The logarithmic likelihood is defined as $\ln \mathcal{L} = -0.5 \sum_{i} \left[ (I_i - I_{\mathrm{mod},i})^2/\delta_i^2 + \ln(2 \pi \delta_i^2) \right]$ with $\delta_i^2 = \delta_\mathrm{int}^2 + \delta I_i^2$. 
The first model is an exponential disc with free scale length $R_\mathrm{d}$ and central surface brightness $I_\mathrm{d,0}$; the second adds a S\'{e}rsic profile with three more parameters, the effective radius $R_\mathrm{e}$, the index $n$, and the central surface brightness $I_\mathrm{b,0}$. 
Prior to the fitting, the models are convolved with a Gaussian PSF with FWHM~$=1\arcsec$. 
To ensure that the model reproduces $I(R)$ at large $R$, we exclude local bumps, possibly \sii\ contamination from spiral arms and star forming regions. 
Priors are set uniform and within reasonable boundaries: $0.4 \arcsec < R_\mathrm{d} < 20 \arcsec$, $0.4 \arcsec < R_\mathrm{e} < f_{R_\mathrm{e}} R_\mathrm{d}$ with $0 < f_{R_\mathrm{e}} < 1$, $0.6 < n < 10$, $15 < I_\mathrm{d,0} < 21.5$ \citep{2011vanderKruitFreeman}, and $15 < I_\mathrm{b,0} < 24 $. 
The lower bound on the scale lengths is set slightly below the PSF effective radius to avoid forcing the fit. 
We compute the Bayesian evidences of the two models ($Z_1$ and $Z_2$) and prefer the second one when $\Delta \ln Z = \ln Z_2 - \ln Z_1 > 2.5$, indicating strong evidence of a bulge component. 
We find evidence for a bulge in $54\%$ of the sample. 
The bulge is resolved ($|R_\mathrm{e} - 0.5\arcsec| > \delta_{R_\mathrm{e}}$) for about $83\%$ of bulge hosts, marginally resolved ($|R_\mathrm{e} - 0.5\arcsec| \approx \delta_{R_\mathrm{e}}$) for $15\%$, and unresolved ($R_\mathrm{e} < 0.5\arcsec$) for A3128\_B\_0248. 
In the \citetalias{myGASPOnlineAtlas}, panel B shows $I(R)$ models.

We then fit the preferred model to $\Sigma_\star(R)$ with \textsc{dynesty}, keeping $R_\mathrm{d}$, $R_\mathrm{e}$, and $n$ fixed and fitting the mass-to-light ratios of disc ($\Upsilon_d^I$) and bulge ($\Upsilon_b^I$), from which we obtain their central surface densities. 
We impose uniform priors within the typical values from stellar population models \citep[e.g.][]{2004Portinari,2022Schombert}: $0.5 < \Upsilon_d^I < 2.5$ and $0.5 < \Upsilon_b^I < 10$. 
The posterior distributions are well constrained for both $\Upsilon_d^I$ and $\Upsilon_b^I$, with few exceptions. 
The medians are $\Upsilon_d^I = 1.1_{-0.3}^{+0.5}$ and $\Upsilon_b^I = 2.0_{-0.8}^{+1.5}$, with $\Upsilon_d^I / \Upsilon_b^I = 0.7_{-0.3}^{+0.1}$. 
We thus assume $\Upsilon^I$ to differ between disc and bulge but to be radially constant in each. 
Since $\Upsilon^I$ depends mostly on stellar age \citep[e.g.][]{2011Tortora}, this is reasonable for bulges, dominated by evolved populations, but may oversimplify discs, which can have non negligible age gradients \citep[e.g.][]{2025Varasteanu}. 
As a sanity check, we convert the model $\Sigma_\star(R)$ into surface brightness using the best-fit $\Upsilon_d^I$ and $\Upsilon_b^I$. 
The agreement with the observed $I(R)$ (panel B in \citetalias{myGASPOnlineAtlas}) is good, with median residuals of $\approx0.1$~mag~arcsec$^{-2}$, so $\Upsilon^I$ is approximately constant within the area probed by our data. About $20$\% of the sample shows a radial trend of $\approx0.2$~mag~arcsec$^{-2}$ per $R_\mathrm{d}$, as expected if the outer disc is younger, which would shift $j_\star$ by $\lesssim0.08$~dex with no dependence on environment or degree of stripping.

\subsection{Tilted-ring modelling: \barolo setup}\label{app:kin_setup}
This section details the configuration adopted for the \barolo\ runs of Sects.~\ref{sec:StarKin} and~\ref{sec:GasKin}. 
For the stellar velocity fields we perform three \texttt{2DFIT} runs per galaxy, progressively fixing more parameters. 
The first run fixes $x_0$, $y_0$, and $i$ at the photometric values and leaves $V_\mathrm{rot,\star}$, $\phi$, and $V_\mathrm{sys}$ free; the second run fixes $V_\mathrm{sys}$ at the median of the first; the third run, our fiducial model, leaves $V_\mathrm{rot,\star}$ as the only free parameter, with $\phi$ fixed at the median of the second run or regularised with a linear model. 
All runs fit both sides (\texttt{side=B}) and use \texttt{wfunc=2}, weighting the major axis, where the rotation signal is strongest. 
When running \texttt{3DFIT} on emission line datacubes, we mask the datacube using \texttt{mask=smooth\&search} and \texttt{factor=2}. 
This smooths the cube with a 2\arcsec\ Gaussian, keeps only pixels with $S/N>5$ (\texttt{snrcut=5}), and grows the mask to adjacent spaxels with $S/N>3$ (\texttt{growthcut=3}).
The MUSE spectral resolution and sampling are $\approx 2.6$~\AA{} ($\approx 110$~\kms) and $\approx 1.25$~\AA{} ($\approx 50$~\kms), so the LSF covers approximately two channels and we set the spectral broadening parameter to \texttt{linear=0.85}.
Regions contaminated by AGN-driven outflows are excluded by setting \texttt{startrad} to the innermost uncontaminated radius. 
A floor of 10~\kms\ is imposed on $\sigma_\mathrm{\hii}$ using \texttt{minvdisp=10}~\kms\ to avoid unrealistic values below the ionised gas thermal velocity. 
We fit both sides simultaneously (\texttt{side=b}), unless bright and kinematically incoherent clumps make it preferable to model only the approaching (\texttt{side=a}) or the receding (\texttt{side=r}) side.
We adopt \texttt{wfunc=2} and \texttt{ftype=2}, the latter minimising the absolute residuals between data and model, normalised to the azimuthally-averaged flux in each ring (\texttt{norm=azim}).
The \texttt{twostage} option runs a second fit in which the geometrical parameters from the first are regularised using a constant or a polynomial. 

\barolo\ assumes a Gaussian PSF, whereas the MUSE one is a Moffat with $\beta\simeq2.5$ \citep{2013Kamann,2016Hodge}, whose wings are much broader at the same FWHM. 
Beam smearing is therefore stronger than \barolo\ assumes and may be under-corrected. 
The MUSE LSF is also more squared than a Gaussian, but the Gaussian approximation is good \citep{2017Bacon} and the effect is minor compared with the spectral resolution.
We estimate these systematic biases to be within the measurement uncertainties and, as for the beam smearing affecting the stellar kinematics maps, independent of environment.
To quantify the effect of beam smearing on the $V_\mathrm{circ}$ inferred from stellar kinematics, we repeat the \halpha\ analysis with the 2D approach used for the stars, running \texttt{2DFIT} on the velocity field with the rings, geometry and datacube of the fiducial \texttt{3DFIT} and measuring $\sigma_\mathrm{\hii}$ from the second-moment map.
$V_\mathrm{rot,\hii}$ is underestimated by $11\%$ between one and two resolution elements and by $4\%$ beyond three, but $\sigma_\mathrm{\hii}$ and hence $V_\mathrm{AD}$ are overestimated, so $V_\mathrm{circ}$ is recovered to within $6\%$ beyond one resolution element, with no clear trend with $i$ or $M_\star$.

To test whether the tilted-ring is missing significant motions other than circular rotation, we also fit the stellar velocity fields using the software \textsc{Kinemetry} \citep{2006Krajnovic} for harmonic decomposition, including terms up to $n=5$. 
We find no clear evidence for multiple kinematic components other than the bars identified by \citet{2023SanchezGarcia} and counter-rotating central components \citep[see][]{2018Moretti_b,2018Vulcani}, and the first-order term agrees with our $V_\mathrm{rot,\star}\sin i$. 
The amplitudes of terms $n>1$ are small compared to the first-order one and become consistent with zero within the uncertainties in the outer disc, where the S/N is lowest.

\subsection{Scaling relations}\label{app:orto_lin_reg}
We perform an orthogonal linear regression using \textsc{dynesty} for nested sampling. 
We shift the coordinate origin to the medians $\overline{x}$ and $\overline{y}$ of the fitted sample (e.g. cluster plus field in Fig.~\ref{fig:TFR}), reducing the parameter covariance. 
The logarithmic likelihood is written to minimise the orthogonal distance between the data points and the best-fit relation, $\ln \mathcal{L} = -0.5 \sum_{i} \left[ d_i^2/\delta_i^2 + \ln(2 \pi \delta_i^2) \right]$, where $d_i = x_i \cos \theta + y_i \sin \theta - \gamma$ is the distance of the $i$-th point from the line, $\theta$ is the perpendicular line angle with the $x$-axis, and $\delta_i^2 = \delta_\perp^2 + (\delta x_i \cos \theta)^2 + (\delta y_i \sin \theta)^2$ combines $\delta_\perp$ with the observational uncertainties $\delta x$ and $\delta y$. 
The slope and intercept follow as $\alpha = - 1/ \tan \theta$ and $\beta = \gamma/\sin \theta + \overline{y} - \alpha \overline{x}$. 
The fitted parameters are thus $\theta$, $\gamma$, and the orthogonal intrinsic scatter $\delta_\perp$, all with uniform priors. 
We resample their posteriors to obtain $\alpha$, $\beta$, and the vertical scatter $\delta_y$ in Table~\ref{tab:scaling_bestfit}. 
The procedure assumes Gaussian $x$ and $y$ uncertainties, while in log space they are not; with the routine \texttt{scattr}\footnote{\url{https://github.com/lucadimascolo/scattr}} \citep[see also][]{2024Rizzo}, which handles non-Gaussian errors, we verify that our conclusions are unaffected.

To quantify the difference between two sub-samples A and B beyond the mere comparison of the best-fit parameters, we measure for each galaxy the vertical residual from the relation fitted to the whole sample, $\Delta_i = y_i - (\alpha x_i + \beta)$, and compare the mean residuals of two sub-samples, $\Delta_y = \langle\Delta\rangle_\mathrm{A} - \langle\Delta\rangle_\mathrm{B}$.
We obtain the uncertainty $\delta_\Delta$ and the significance by reshuffling the sub-sample labels $2\times10^4$ times at fixed sub-sample size: if the environment plays no role the labels carry no information, and the permutations give the distribution of $\Delta_y$ produced by chance alone at our sample size and scatter. 
We take $\delta_\Delta$ as the standard deviation of that distribution and $p_\mathrm{null}$ as the fraction of reshufflings giving an offset at least as large as the observed one, that is the probability of measuring it by chance if the environment makes no difference.
We repeat the test on the common range of $x$ of the two sub-samples ($V_\mathrm{flat}$ for the TFRs and $M$ for the SAMRs), when A and B do not span the same range.

\subsection{Comparison with HI-based Tully-Fisher relations}\label{app:M25}
The stellar TFR of \citet{2025Marasco} for the SPARC sample is much steeper ($\alpha = 5.21 \pm 0.18$) than ours, despite the similar methods used to measure the kinematics, $V_\mathrm{flat}$, and $M_\star$ (Sect.~\ref{sec:literature_TFR}). This discrepancy has two causes. 
The first is the different $M_\star$ range.
Applying our fitting routine to the \citet{2025Marasco} sample, we find essentially the same best fit as they do ($\alpha = 5.11_{-0.13}^{+0.14}$, $\beta = -0.96_{-0.30}^{+0.28}$, and $\delta_y = 0.25 \pm 0.03$) over the full range $7 \lesssim \log (M_\star/\mathrm{M_\odot}) \lesssim 11.5$, but we obtain $\alpha = 4.60_{-0.18}^{+0.18}$, $\beta = 0.21_{-0.41}^{+0.40}$, and $\delta_y = 0.17 \pm 0.03$ when restricting to the same range $9 \lesssim \log (M_\star/\mathrm{M_\odot}) \lesssim 11.5$ as our sample.
Our TFRs are thus biased towards shallower slopes partly because we have few or no galaxies below $\log (M_\star/\mathrm{M_\odot}) \approx 9.5$. 
The second cause is the rotation curve extent: \cite{2019Lelli} give empirical relations linking $V_\mathrm{flat,\hi}$ to other velocity definitions ($V_x$),
\begin{equation}\label{eq:VflatHItoVflatOther}
	\log V_\mathrm{flat,\hi} = \eta \log V_x + \zeta \,,
\end{equation}
with $\eta \approx 0.8$, $\zeta \approx 0.4-0.5$, and scatter $\delta_y \approx 0.034-0.037$ when $V_x$ is measured at $2.2 R_\mathrm{d}$ or $2 R_\mathrm{e}$, roughly where we measure $V_\mathrm{flat}$.
Substituting Eq.~\ref{eq:VflatHItoVflatOther} in Eq.~\ref{eq:TFR}, a TFR based on $V_x$ has slope $\alpha = \alpha' \eta$ and normalisation $\beta = \alpha' \zeta + \beta'$, where $\alpha'$ and $\beta'$ are the parameters of the relation based on $V_\mathrm{flat,\hi}$.
Applying these transformations to the high-$M_\star$ fit on the \citet{2025Marasco} sample, we predict $\alpha \approx 3.58$, $\beta \approx 2.48$, and $\delta_y \approx 0.24$ for a TFR based on $V(2.2R_\mathrm{d})$, in excellent agreement with our best fit ($\alpha \approx 3.58$, $\beta \approx 2.43$, $\delta_y \approx 0.21$).
The different $M_\star$ range and rotation curve extent thus fully reconcile the values derived here with those in \citet{2025Marasco}.

\section{More on the impact of ram pressure and environment}\label{app:gas_vs_JType}
This section looks more closely at the effect of ram pressure on the gas kinematics, distribution, and $j$. 

\subsection{Rotation-to-dispersion ratio}\label{app:Vsigma_rps}
\begin{figure}
	\centering
	\includegraphics[width=1\linewidth]{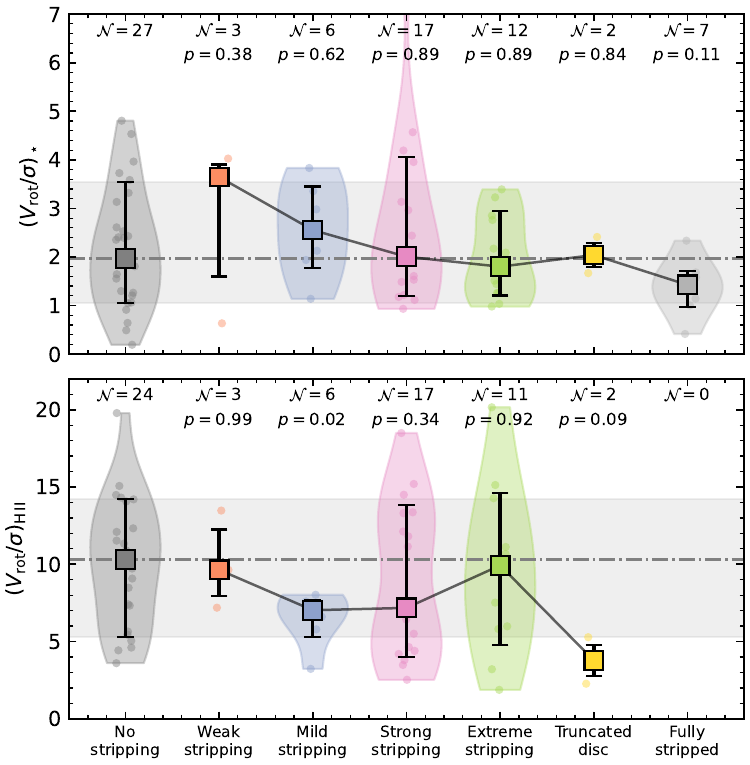}
	\caption{Stellar (top) and ionised gas (bottom) $V_\mathrm{rot}/\sigma$ as a function of degree of ram pressure stripping. 
			The violins show $V_\mathrm{rot}/\sigma(R)$ distribution for each Jtype.
			Shaded dots are the median $V_\mathrm{rot}/\sigma$ for individual galaxies, with the squares being the median for each JType. 
			The grey dash-dotted line and shaded band indicate the median and the 16th-84th percentile range of the non-stripped sample, respectively. 
			$\mathcal{N}$ is the number of galaxies in each class and $p$ the KS null probability against the non-stripped sample.}
	\label{fig:V_over_sigma_RPS}
\end{figure}

Figure~\ref{fig:V_over_sigma_RPS} compares $(V_\mathrm{rot}/\sigma)_\star$ and $(V_\mathrm{rot}/\sigma)_\mathrm{\hii}$ of galaxies currently, or formerly for truncated discs, affected by ram pressure against unperturbed ones. 
Both panels show mild but visible decreasing trends with JType, whose significance we assess with KS and Spearman tests, pooling all stripping stages against the non-stripped sample for statistics; treating each JType separately does not change our conclusions.
For $(V_\mathrm{rot}/\sigma)_\star$ the Spearman coefficient is $\rho_\mathrm{Spear} \approx -0.08$ ($p_\mathrm{Spear} \approx 0.49$), and the KS test gives $p_\mathrm{null}\gtrsim0.87$: the stripped and unperturbed distributions are statistically indistinguishable, as expected since ram pressure does not act on the stars. 
$(V_\mathrm{rot}/\sigma)_\mathrm{\hii}$ instead decreases marginally with JType ($\rho_\mathrm{Spear}=-0.20$, $p_\mathrm{Spear} \approx 0.10$), with $p_\mathrm{null} \approx 0.15$ that the two distributions share a parent. 
Only the mild stripping cases have $p_\mathrm{null}<0.05$, but there is no ordered trend for the other JType classes.
These results are not statistically firm: excluding the rings with $\sigma_\mathrm{\hii}<30$~\kms, which are upper limits, removes the high-$\sigma_\mathrm{\hii}$ tail of the undisturbed discs, reverses the marginal decrease with JType, and lowers the KS probability to $p_\mathrm{null}\approx 0.05$. 
Any correlation of $(V_\mathrm{rot}/\sigma)_\mathrm{\hii}$ with JType is therefore secondary, or undetectable with our methods. 

\begin{figure*}
	\centering
	\includegraphics[width=1\textwidth]{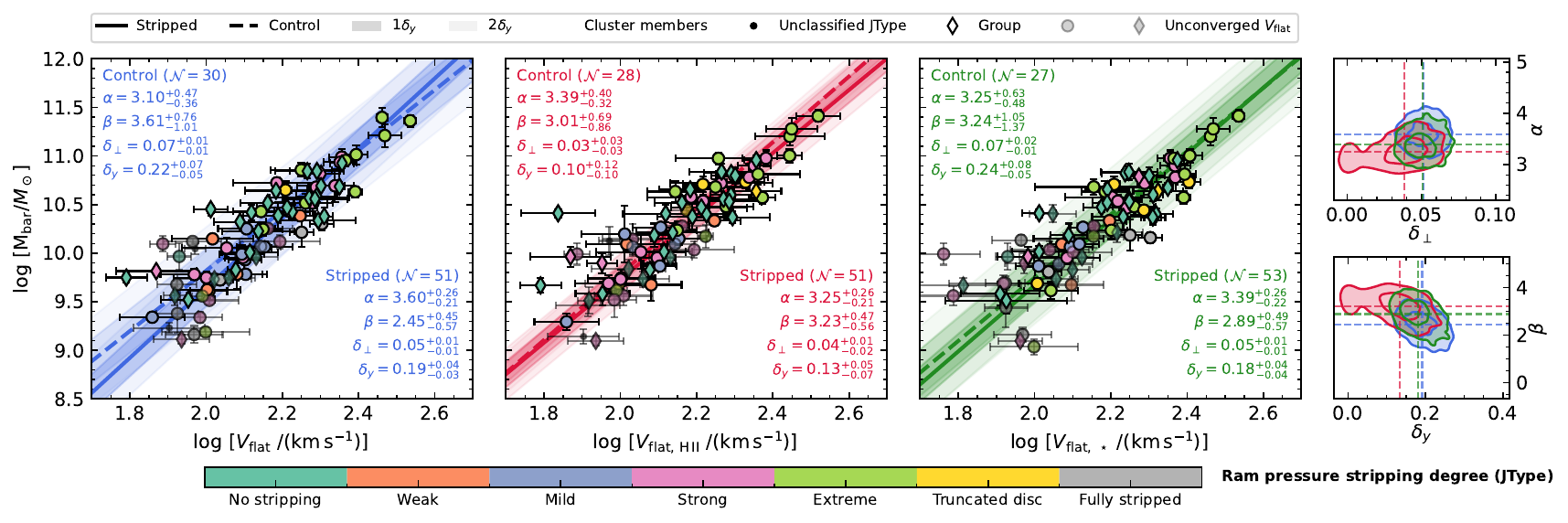}
	\caption{Baryonic TFR coloured by degree of ram pressure stripping, with $V_\mathrm{flat}$ from the fiducial $V_\mathrm{circ}$ (left), the ionised gas (centre), and the stars (right). 
		Solid and dashed lines are the best fits for the stripped (JType$>0$) and control (JType$\leq0$) samples, circles and diamonds are cluster and field galaxies. 
		The small panels on the right compare the 2D posteriors of the three tracers for stripped galaxies: they overlap fully or within $\approx 2 \delta$, indicating no significant difference between tracers, as found for the two samples. }
	\label{fig:TFR_JType}
\end{figure*}

\subsection{TFRs with different tracers}\label{sec:TFR_rps}
We now explore whether the position of galaxies on the TFRs depends on the degree of ram pressure stripping (JType; Sect.~\ref{sec:class} and Table~\ref{tab:sample}). 
We focus on the baryonic TFR for cluster galaxies, where the ram pressure might affect both $V_\mathrm{flat}$ (perturbing the gas kinematics) and $M_\mathrm{bar}$ (removing gas). 
The left panel of Fig.~\ref{fig:TFR_JType} shows the baryonic TFR from our fiducial $V_\mathrm{flat}$: there is no clear offset with the degree of ram pressure stripping. 
Galaxies undergoing strong or extreme stripping follow the same relation as the others, despite being expected to have the largest kinematic perturbation. 
The same holds for group members. 

The central and right panels of Fig.~\ref{fig:TFR_JType} show that the result does not depend on the kinematic tracer: even strongly and extremely stripped galaxies follow the same TFR whether $V_\mathrm{flat}$ is from the ionised gas (central panel) or the stars (right panel), which are not perturbed by ram pressure. 
The three sets of 2D posteriors largely overlap, the gas giving a marginally shallower and tighter relation because $\delta V_\mathrm{rot, \hii} > \delta V_\mathrm{rot, \star}$. 
This is unsurprising, since gas and stellar $V_\mathrm{circ}$ are very similar, with few exceptions. 
An instructive example is JW100, a jellyfish galaxy with very extended tails of ionised gas: despite the ongoing ram pressure, its gas and stellar $V_\mathrm{circ}(R)$ agree within the uncertainties, except for the central 3-4~kpc where the ionised gas is strongly contaminated by the AGN outflow. 
The lack of any segregation with JType suggests that, within the disc area analysed here, any change in the baryonic kinematics and gravitational potential is too small to produce significant deviations, even for truncated and fully stripped discs (S0 and S0/a). 

\subsection{$j$ and phase-space location}\label{sec:AMR_phasespace}
The phase-space diagram shows the LOS velocity of a galaxy ($V_\mathrm{gal}$) normalised to the cluster velocity dispersion ($\sigma_\mathrm{cl}$) against the galaxy cluster-centric distance ($d_\mathrm{gal}$) normalised to the cluster virial radius ($R_\mathrm{vir}$). 
Galaxies preferentially occupy different phase-space regions depending on their infall time ($t_\mathrm{inf}$). 
Stronger environmental effects are expected for earlier $t_\mathrm{inf}$, and the ram pressure strength is expected to peak at small $d_\mathrm{gal}$ (high ICM density) and large $V_\mathrm{gal}$. 
The phase-space diagram might thus reveal $j$ trends not seen in Figs.~\ref{fig:AMR} and~\ref{fig:AMR_JType}. 

\begin{figure}[h]
	\centering
	\includegraphics[width=1\columnwidth]{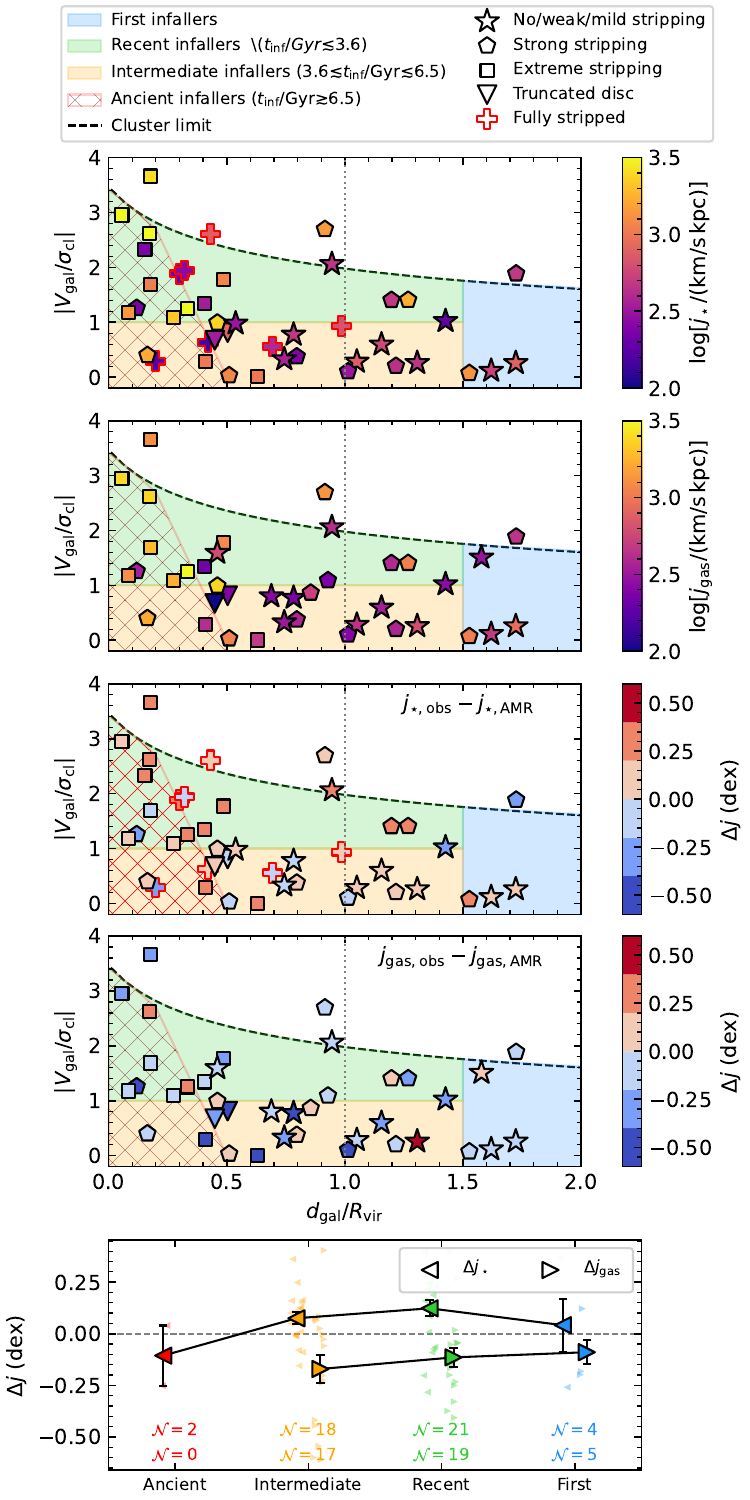}
	\caption{Phase-space diagram with galaxies coloured for $j_\star$ (first panel), $j_\mathrm{gas}$ (second), and vertical offset $\Delta j$ from the stellar (third) and gas (fourth) SAMR for field galaxies. 
		Coloured areas are zones where it is most likely to find galaxies with given infall time ($t_\mathrm{inf}$) range. 
		The bottom panel shows $\Delta j$ vs $t_\mathrm{inf}$: galaxies are assigned $t_\mathrm{inf}$ based on their closest zone or, when two overlap, fully stripped, truncated, and extreme-to-mild stripping are assumed ancient, intermediate, and recent infallers, respectively. } 
	\label{fig:phase_space}
\end{figure}

Figure~\ref{fig:phase_space} shows the position of our galaxies in phase-space, with $d_\mathrm{gal}$ and $\sigma_\mathrm{cl}$ from \cite{2017Biviano} and \cite{2020Gullieuszik}. 
The shaded areas mark where galaxies with a given range of $t_\mathrm{inf}$ are most likely found, from the simulations of \cite{2017Rhee} (their Fig.~9, for massive sub-halos). 
These areas are indicative, not sharp boundaries, since $d_\mathrm{gal}$ and $V_\mathrm{gal}$ are projected on the sky and both interloper contamination and the dispersion in $t_\mathrm{inf}$ are significant \citep[e.g.][]{2025Dou}. 
Galaxies with high $j_\star$ (first panel) and $j_\mathrm{gas}$ (second panel) are mostly jellyfish and predominantly at small $d_\mathrm{gal}/R_\mathrm{vir}$, as expected for recent infallers \citep[see also][]{2018Jaffe}; $j_\star$ and $j_\mathrm{gas}$ do not appear to correlate with $t_\mathrm{inf}$. 
The third and fourth panels colour cluster members by their vertical offset $\Delta j$ from the field stellar and gas SAMRs. 
$\Delta j_\star$ does not correlate with phase-space position or $t_\mathrm{inf}$ (fifth panel), except perhaps for the jellyfish galaxies, the most recent infallers \citep{2018Jaffe}, which have $\Delta j_\star \approx +0.2$~dex. 
$\Delta j_\mathrm{gas}$ appears uncorrelated too: intermediate, recent, and first infallers give $-0.17\pm0.07$, $-0.11\pm0.05$, and $-0.09\pm0.06$~dex, consistent within the uncertainties, and no galaxy with $j_\mathrm{gas}$ falls in the ancient infaller region. 
Fully stripped discs (S0 and S0/a) sit mostly in the ancient and intermediate infaller regions \citep[see also][]{2020Vulcani_a}, but their $j_\star$ does not correlate with the phase-space location either. 

\subsection{Toy model for stripped discs}\label{app:toy}
\begin{figure}
	\centering
	\includegraphics[width=0.5\textwidth]{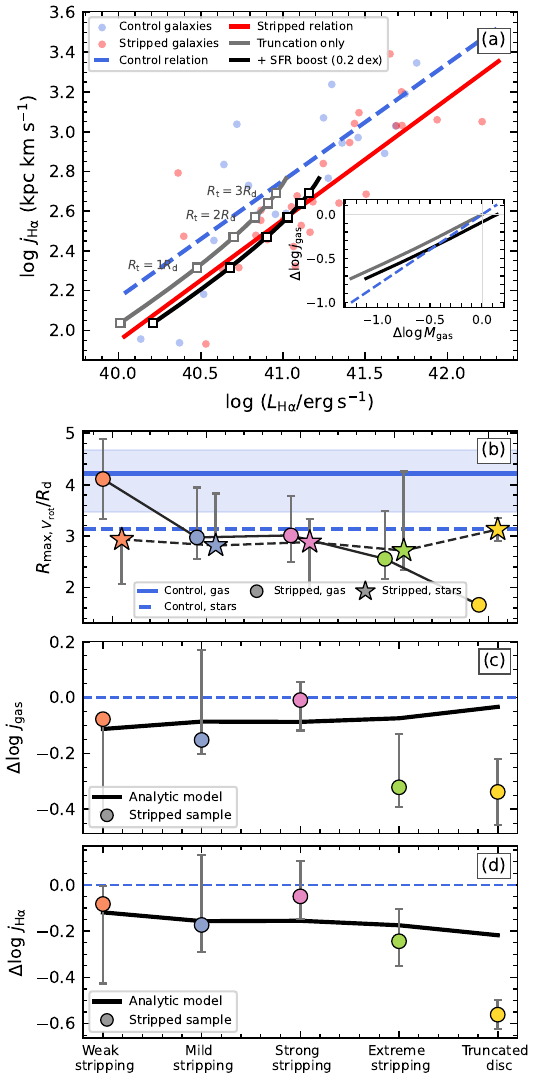}
	\caption{
	\emph{Panel (a)}: $j_\mathrm{\halpha}$--$L_\mathrm{\halpha}$ plane. Points are the control (blue) and stripped (red) galaxies, lines their best-fit relations. The grey and black curves are the tracks of a galaxy starting on the control relation as its disc is truncated, without and with the SFR boost; squares mark various $R_\mathrm{t}/R_\mathrm{d}$. \emph{Inset}: the same tracks in the plane of the gas offsets.
	\emph{Panel (b)}: outermost radius with measured $V_\mathrm{rot}$ for gas and stars, normalised to the disc scale length, for each JType class (medians and 16th--84th percentiles) and for the control sample (blue lines and band).
	\emph{Panels (c) and (d)}: offsets of the stripped galaxies from the control relations, per JType class, in the gas and \halpha\ planes. The black curve is the analytic model, with the truncations of panel (b); symbols are the measured medians with 16th--84th percentiles, after subtracting the median residual of the control galaxies.}
	\label{fig:toymodel}
\end{figure}

We interpret the negative offset of the stripped galaxies from the control gas SAMR with a toy model (Fig.~\ref{fig:toymodel}). 
An exponential \halpha\ disc of scale length $R_\mathrm{d}$ and flat circular velocity $V_\mathrm{circ}$ has, within $x = R_\mathrm{t}/R_\mathrm{d}$,
\begin{equation*}
	j(x) = V_\mathrm{circ} R_\mathrm{d}\,\frac{g(x)}{f(x)} \xrightarrow[x\to\infty]{} 2 V_\mathrm{circ} R_\mathrm{d} \,,
\end{equation*}
where $f(x)=\int_0^x t\,e^{-t}dt = 1-e^{-x}(1+x)$ and $g(x)=\int_0^x t^2e^{-t}dt = 2-e^{-x}(x^2+2x+2)$ are the cumulative mass and angular momentum of the disc.
Assuming that $\Sigma_\mathrm{gas}(R)$ has the same radial gradient as $\Sigma_\mathrm{SFR}(R)$ (Sect.~\ref{sec:GasDist}), we have $j_\mathrm{gas}=j_\mathrm{\halpha}$, $M_\mathrm{gas}=\pi R_\mathrm{max}^2\langle\Sigma_\mathrm{gas}\rangle$, and
\begin{equation*}
	\log M_\mathrm{gas} = \frac{1}{N}\log L_\mathrm{\halpha}
	+ 2\Big(1-\frac{1}{N}\Big)\log R_\mathrm{max} + \mathrm{const},
\end{equation*}
with $N=1.41$ the SK law index. 

A galaxy starts on the control relation with $x_\mathrm{ref}=4.2$, the median $R_\mathrm{max,V_\mathrm{rot}}/R_\mathrm{d}$ of the control sample.
Ram pressure erodes the disc to $x_\mathrm{t}<x_\mathrm{ref}$ and enhances the SFR by $\mathcal{B}\simeq0.2$~dex \citep{2018Vulcani_b}, so that
\begin{gather*}
	\Delta\log L_\mathrm{\halpha} = \log\frac{f(x_\mathrm{t})}{f(x_\mathrm{ref})}+\mathcal{B},
	\quad
	\Delta\log j = \log \frac{j(x_\mathrm{t})}{j(x_\mathrm{ref})}, \\
	\Delta\log M_\mathrm{gas} = \frac{\Delta\log L_\mathrm{\halpha}}{N}
	+ 2\Big(1-\frac{1}{N}\Big)\log\frac{x_\mathrm{t}}{x_\mathrm{ref}}.
\end{gather*}
The boost $\mathcal{B}$ raises $\Sigma_\mathrm{SFR}$ everywhere by the same factor, leaving $\Delta\log j$ unchanged.
The offsets from the control relations, of slopes $\alpha_\mathrm{\halpha}$ and $\alpha_\mathrm{gas}$, are then
\begin{equation*}
	\Delta_\mathrm{\halpha} = \Delta\log j - \alpha_\mathrm{\halpha}\Delta\log L_\mathrm{\halpha},
	\quad
	\Delta_\mathrm{gas} = \Delta\log j - \alpha_\mathrm{gas}\Delta\log M_\mathrm{gas}.
\end{equation*}
Since $\alpha_\mathrm{gas}\simeq N\alpha_\mathrm{\halpha}$, the two are related by $\Delta_\mathrm{gas}\simeq\Delta_\mathrm{\halpha}-2\alpha_\mathrm{gas}(1-N^{-1})\log(x_\mathrm{t}/x_\mathrm{ref})$ (verified numerically). 

For the observed median truncation of the stripped sample, $x_\mathrm{t}=2.9$, the model gives $\Delta_\mathrm{\halpha}=-0.16$ and $\Delta_\mathrm{gas}=-0.09$~dex, against the measured $-0.13$ and $-0.11$~dex.
Truncation alone would give only $-0.04$ and $+0.03$~dex, hence the offset comes from the SFR enhancement, which raises $L_\mathrm{\halpha}$ (and thus the $M_\mathrm{gas}$ inferred through the SK law) without changing $j$. 
The model also predicts saturation: the offset reaches $-0.25$~dex at $x_\mathrm{t}\simeq0.7$ and becomes less negative for stronger truncation, because a small residual disc loses mass faster than $j$.
Over the observed range of truncations, the predicted offset varies little with JType (Fig.~\ref{fig:toymodel}), which is what we measure, and explains why $\Delta j_\mathrm{\halpha}$ does not correlate with $R_\mathrm{max}/R_\mathrm{d}$ among the stripped galaxies ($\rho_\mathrm{Spear}=+0.21$, $p_\mathrm{Spear}=0.23$).
Only the two truncated discs lie beyond the saturation floor.

We also calculate how much the rotation curve could have changed to produce the offset. 
If $V_\mathrm{circ}$ is reduced by a factor $1-\varepsilon$ for $R>\xi R_\mathrm{d}$, $L_\mathrm{\halpha}$ and $M_\mathrm{gas}$ are unchanged, so the offset is entirely due to $j$:
\begin{equation*}
	\Delta\log j = \log\left[1-\varepsilon\,W(\xi)\right] \,,
\end{equation*}
with $W(\xi)=[g(x_\mathrm{t})-g(\xi)]/g(x_\mathrm{t})<1$ the fraction of the angular momentum contributed at $R>\xi R_\mathrm{d}$.
Since $V_\mathrm{flat}$ comes from the outermost points of $V_\mathrm{circ}(R)$ (Sect.~\ref{sec:TFR}), it is reduced by $\Delta\log V_\mathrm{flat}=\log(1-\varepsilon)$, whereas $j$ is reduced less, since $W(\xi)<1$.
A change in the rotation curve thus always leaves a larger imprint on the TFR than on the SAMR, because $|\Delta\log j| \le |\Delta\log V_\mathrm{flat}|$ for any $\xi$.
Since we detect no offset from the TFRs (Sect.~\ref{sec:literature_AMR}), where $\Delta_y$ is within $\delta_y$ and thus $V_\mathrm{circ}$ cannot be reduced by more than $\approx14$\%, we obtain $|\Delta\log j| \lesssim 0.05$~dex, a third of the observed offset.
Moreover, since $V_\mathrm{circ}=\sqrt{V_\mathrm{rot}^2+V_\mathrm{AD}^2}$, a decrease of $V_\mathrm{rot}$ compensated by a larger $V_\mathrm{AD}$ would leave $V_\mathrm{circ}$ unchanged, and with it $j_\mathrm{gas}$ itself, so it could not produce the observed offset either.
In practice $V_\mathrm{AD}$ is negligible where it matters: at $R>R_\mathrm{d}$, we measure $V_\mathrm{circ}/V_\mathrm{rot}=1.014$ and $1.030$ for control and stripped galaxies, that is $\lesssim0.01$~dex in velocity.
The ratio does grow mildly with JType ($\rho_\mathrm{Spear}=+0.38$, $p_\mathrm{Spear}=0.04$), consistent with the increase of pressure support in Fig.~\ref{fig:V_over_sigma_RPS}, but by at most $0.02$~dex, an order of magnitude too little to hide the velocity change required to produce the offset.

\section{More on $j$ computation}\label{app:more_j}
This section contains some details on the computation of $j$. 

\subsection{Convergence}\label{app:j_convergence}
\begin{figure}
	\centering
	\includegraphics[width=1\columnwidth]{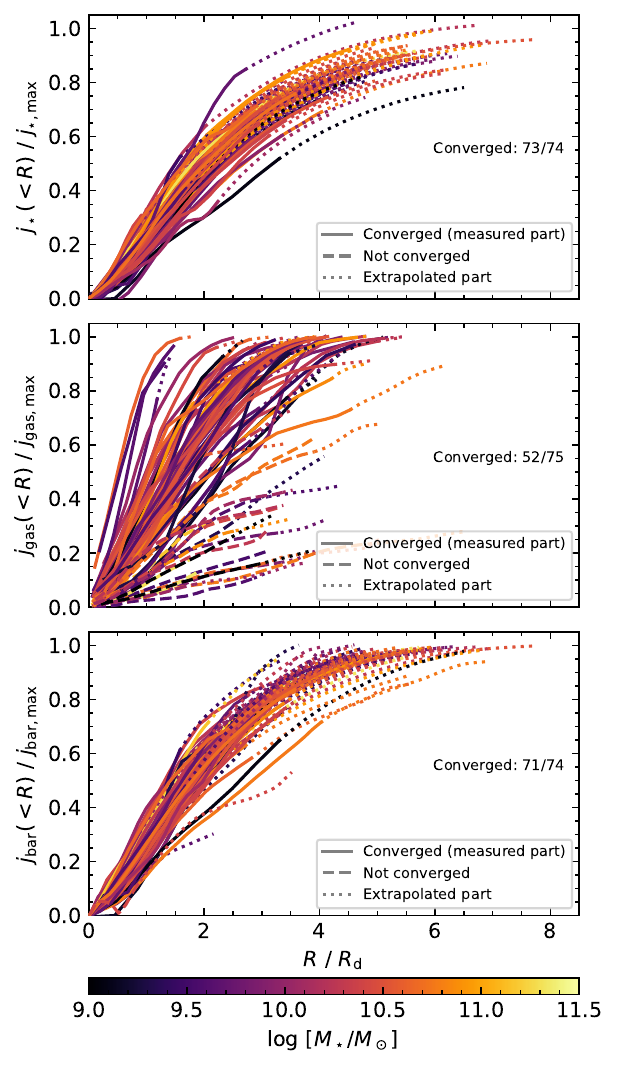}
	\caption{Cumulative profiles of $j_\star$, $j_\mathrm{gas}$, and $j_\mathrm{bar}$, coloured by $M_\star$. 
		Dotted segments are extrapolated to $R_\mathrm{max}$; solid and dashed curves are converged and non-converged profiles.}
	\label{fig:cumj_profiles}
\end{figure}

Figure~\ref{fig:cumj_profiles} shows $j(R)$ for each baryonic component in the disc, distinguishing between galaxies meeting or not the convergence criterion of \citet{2021ManceraPina}. 
We fit a second order polynomial to the outer 30\% of $j(R)$. 
The profile is considered converged if its outermost point exceeds 80\% of the polynomial asymptotic value; thus, we recover the total $j$ within $\approx 15$\% ($\delta j<0.1$~dex; \citealt{2021ManceraPina}).
Extrapolating to $R_\mathrm{max}$ makes $j_\star(R)$ and $j_\mathrm{bar}(R)$ converge in most cases, since $R_\mathrm{max} > 2 R_\mathrm{d}$ and stars dominate the baryons; $j_\mathrm{gas}(R)$ converges instead for $\approx 70$\% of the galaxies. 
Including galaxies with non-converged $j$ in the SAMR analysis leaves the best-fit slopes unchanged, while the offset between stripped and control galaxies in the gas SAMR is reduced to $-0.10\pm0.06$~dex, preserving its sign and direction. 

\subsection{$j_\mathrm{H\alpha}$-$L_\mathrm{H\alpha}$ relations}\label{app:jhalpha}
\begin{figure*}
	\centering
	\includegraphics[width=1\linewidth]{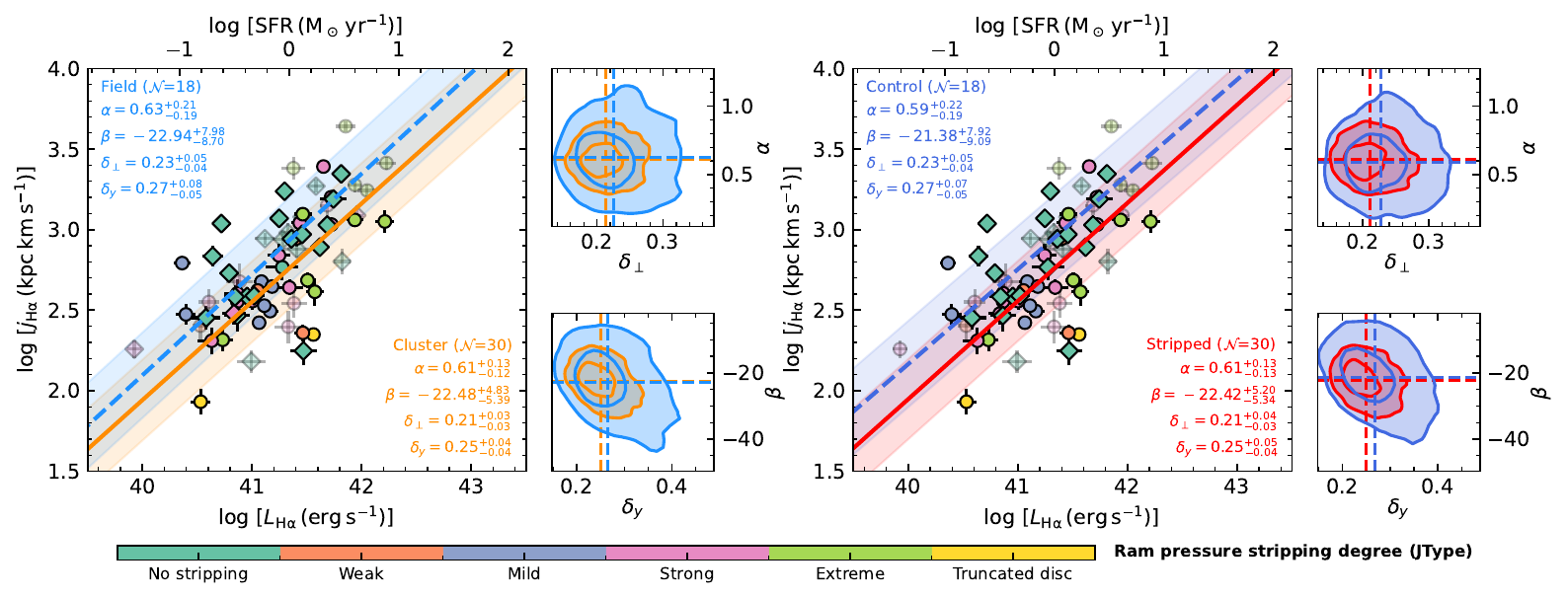}
	\caption{Correlation between $j_\mathrm{H\alpha}$ and $L_\mathrm{H\alpha}$. 
		Left and right panels compare the relation for cluster (solid orange) vs field (dashed blue) and control (dashed blue) vs stripped (solid red) galaxies. }
	\label{fig:jhalpha}
\end{figure*}

Figure~\ref{fig:jhalpha} shows the correlation between the integrated \halpha\ luminosity $L_\mathrm{H\alpha}$, also expressed as SFR, and $j_\mathrm{H\alpha}$, computed from Eq.~\ref{eq:j_1comp} with $\Sigma_\mathrm{H\alpha}$ and $V_\mathrm{rot,H\alpha}$, without assuming the SK law. 
Cluster and field galaxies follow the same relation, with parameters consistent within $1\delta$, so the $\alpha$ difference of Sect.~\ref{sec:AMR_env} is not statistically significant and comes mostly from the \halpha-to-gas conversion: at fixed $L_\mathrm{H\alpha}$, cluster members have \halpha\ discs a factor $\approx1.3$ more concentrated than field galaxies ($R_\mathrm{max}/R_\mathrm{d} = 3.1$ against $4.1$) and therefore higher $\Sigma_\mathrm{gas}$ for a super-linear SK law. 
The same holds for control and stripped galaxies, and the $j$ deficit of the latter survives, with no dependence on JType in either plane (Sect.~\ref{sec:AMR_rps}). 

\subsection{Inclusion of bulges}\label{app:j_bulge}
\begin{figure}
	\centering
	\includegraphics[width=1\linewidth]{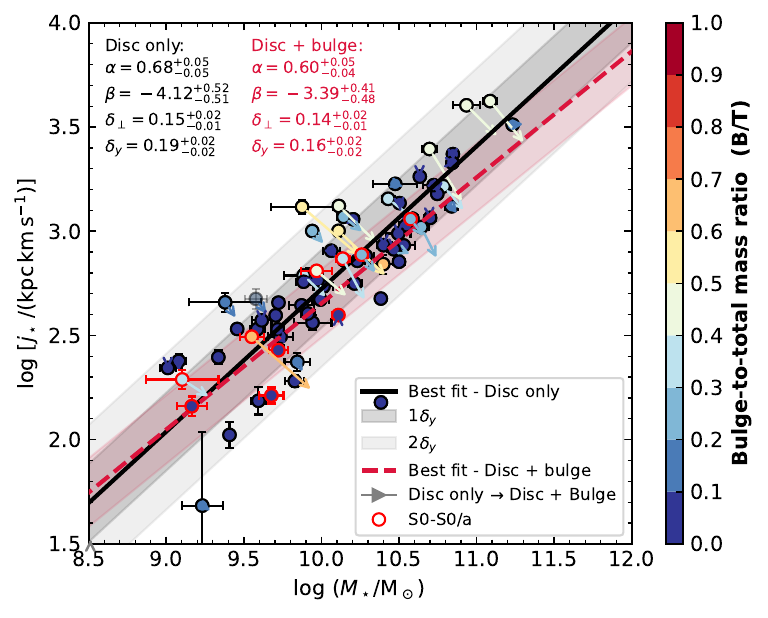}
	\caption{Same as the left panel in Fig.~\ref{fig:AMR} but with galaxies coloured for their B/T. 
		The arrows indicate where galaxies move if the bulge is included in $j_\star$ computation. }
	\label{fig:AMR_BT}
\end{figure}

Figure~\ref{fig:AMR_BT} shows the $j_\star$--$M_\star$ plane including the bulge.
Most galaxies (including S0s and S0/as) are unaffected owing to the low B/T, while some approach the relation. 
The best-fit SAMR remains fully consistent with the one for discs only in Fig.~\ref{fig:AMR} \citep[see also][]{2018Fall}. 
JW105, an S0/a with B/T$\approx 0.6$, is the only galaxy that moves $\gtrsim 2 \delta_y$ from the relation, having $j_\star$ about $0.4$~dex lower. 

\end{appendix}

\end{document}